\documentclass[letterpaper]{JHEP3}
\usepackage{epsfig}

\def\hg{\hat{g}}
\def\hG{\hat{\Gamma}}
\def\hR{\hat{R}}
\def\cR{\check{R}}
\def\exd{{\hbox{d}}}
\def\d{\exd}

\def\Rmu#1#2{{{R^{#1}}}_{{#2}}}
\def\Rml#1{R_{{#1}}}
\def\Rc#1{R_{{#1}}}
\def\hRmu#1#2{{{\hat{R}^{#1}}}_{\;\;{#2}}}
\def\hRml#1{\hat{R}_{{#1}}}
\def\hRc#1{\hat{R}_{{#1}}}

\def\ba{\begin{eqnarray}}
\def\ea{\end{eqnarray}}
\def\be{\begin{equation}}
\def\ee{\end{equation}}
\def\ssB{{\scriptscriptstyle B}}
\def\ssM{{\scriptscriptstyle M}}
\def\ssN{{\scriptscriptstyle N}}
\def\ssP{{\scriptscriptstyle P}}
\def\ssR{{\scriptscriptstyle R}}
\def\ssT{{\scriptscriptstyle T}}
\def\ssW{{\scriptscriptstyle W}}
\def\ssZ{{\scriptscriptstyle Z}}

\def\L{\mathcal{L}}

\def\X{\mathcal{X}}

\def\nn{\nonumber}

\def\d{\mathrm{d}}

\def\({\left(}
\def\){\right)}

\def\hg{\hat{g}}
\def\cg{\check{g}}

\def\pref#1{(\ref{#1})}

\def\eff{{\rm eff}}

\def\ctwoTensn{{T_2}}
\def\ctwoPotl{{U_2}}
\def\coneTensn{{T_1}}
\def\coneKin{{Z_1}}
\def\redtwoTensn{{\cal T}_2}
\def\redtwoPotl{{\cal U}_2}
\def\redoneTensn{{\cal T}}
\def\redoneKin{{\cal Z}}
\def\renctwoTensn{\overline{T}_2}
\def\renctwoPotl{\overline{U}_2}

\title{Effective Field Theories and Matching for Codimension-2 Branes}

\author{C.P. Burgess,${}^{1-3}$ D. Hoover,${}^4$ C. de
Rham${}^{1,2}$ and G. Tasinato${}^5$ \\
${}^1$ Perimeter Institute for Theoretical Physics, Waterloo ON,
N2L 2Y5, Canada.\\
${}^2$
Physics \& Astronomy, McMaster University, Hamilton ON, L8S 4M1,
Canada. \\
${}^3$ Theory Division, CERN, CH-1211 Geneva 23, Switzerland. \\
${}^4$ Physics Dept., McGill University, Montr\'eal, QC, H3A 2T8,
Canada.\\
${}^5$ Institut f\"ur Teoretische Physik, Universit\"at Heidelberg,
D-69120 Heidelberg, Germany
 }

\date{}

\abstract {It is generic for the bulk fields sourced by branes
having codimension two and higher to diverge at the brane
position, much as does the Coulomb potential at the position of
its source charge. This complicates finding the relation between
brane properties and the bulk geometries they source. (These
complications do not arise for codimension-1 sources, such as in
RS geometries, because of the special properties unique to
codimension one.) Understanding these relations is a prerequisite
for phenomenological applications involving higher-codimension
branes. Using codimension-2 branes in extra-dimensional
scalar-tensor theories as an example, we identify the classical
matching conditions that relate the near-brane asymptotic
behaviour of bulk fields to the low-energy effective actions
describing how space-filling codimension-2 branes interact with
the surrounding extra-dimensional bulk. We do so by carefully
regulating the near-brane divergences, and show how these may be
renormalized in a general way. Among the interesting consequences
is a constraint relating the on-brane curvature to its action,
that is the codimension-2 generalization of the well-known
modification of the Friedmann equation for codimension-1 branes.
We argue that its interpretation within an effective field theory
framework in this case is as a relation $4\pi \ctwoPotl \simeq
\kappa^2 \left( {\ctwoTensn}' \right)^2$ between the codimension-2
brane tension, $\ctwoTensn(\phi)$, and its contribution to the
low-energy on-brane effective potential, $\ctwoPotl(\phi)$. This
relation implies that any dynamics that minimizes a brane
contribution to the on-brane curvature automatically also
minimizes its couplings to the extra-dimensional scalar.}

\begin{document}

\section{Introduction}

The study of codimension-1 branes is very well developed, largely
due to the recognition \cite{RS} that the warping induced by
branes can provide new ways to generate hierarchies. Much less is
known about the interactions of higher codimension branes with
their environments. But systems with only one codimension are not
representative of those having more, and the absence of such
studies is likely to strongly bias our understanding of the kinds
of physics to which branes can lead \cite{ButSee}.

There is a good reason why such studies have not been done. The
problem is that (unlike codimension one) for generic codimension
the fields sourced by a brane typically diverge at the brane
position \cite{nonconicalcod2} --- indeed the Coulomb potential
outside a point charge in 3 spatial dimensions provides a familiar
case in point. Such divergences complicate the extraction of
useful consequences of brane-bulk interactions, because these
often require knowing how the properties of the bulk fields are
related to the choices made about the brane-localized physics.
Examples of questions that hinge on this kind of connection arise
in brane cosmology \cite{6Dbranecosmo}, where one wishes to know
how a given energy density and pressure on the brane interacts
with the time-dependent extra-dimensional cosmological spacetime,
or in particle phenomenology \cite{6DHiggs}. The connection is
also crucial for attempts to use extra dimensions to address the
cosmological constant problem
\cite{conicaldefects,SLED,fluxquantc,XDCC,antiSLED}, since these hinge on
understanding the connection between bulk curvatures and radiative
corrections on the brane.

In this paper our goal is to develop tools to remedy this
situation, adapted for studying the fields sourced by
$d$-dimensional space-filling branes sitting within a ($D = d +
2$)-dimensional spacetime. Such codimension-2 branes provide the
simplest possible laboratory to study the problem of how branes
generically interact with their surrounding bulk, and are likely
much more representative of the generic higher-codimension
situation than are the codimension-1 systems presently being
studied. Of special interest is the case where $d=4$, which
describes 3-branes sitting within a 6-dimensional spacetime.

We attain this goal by identifying which features of the bulk
fields are directly dictated by the branes, and showing precisely
how these features depend on the brane action. What we find
resembles what one would expect based on the electrodynamics of
charge distributions situated within an extra-dimensional bulk:
the field behaviour very near a branes is directly governed by
that brane's properties, while overall issues like equilibrium or
stability depend on the global properties of all of the branes
taken together.

More precisely, the success of our analysis relies on there being
a large hierarchy between the small size, $r_b$, of the source
distribution, compared with the large size, $L$, over which the
external field of interest varies. In the electrostatic analogy it
is the existence of distances $r$ satisfying $r_b \ll r \ll L$
that allows the use of a multipole expansion to relate powers of
$r_b/r$ to various moments of the source distribution at distances
much smaller than the scale, $L$. We assume a similar hierarchy
exists in the case of gravitating codimension-2 branes, where
$r_b$ is of order whatever physics governs the branes' microscopic
structure, while $L$ is more characteristic of the curvature or
volume of the geometry transverse to the branes.

Mathematically, we identify the matching conditions that relate
the action of the effective field theory governing the low-energy
properties of the brane with the asymptotic near-brane properties
of the bulk fields they source. These provide the analogue for
higher codimension of the well-known Israel junction conditions
\cite{IJC} that determine the matching of codimension-1 branes to
their adjacent bulk geometries. We derive these conditions by
regularizing the codimension-2 brane by replacing it with an
infinitesimal codimension-1 brane that encircles the position of
the codimension-2 object of interest. This allows the connection
between brane and bulk to be obtained explicitly using standard
jump conditions at this codimension-1 position. We then show how
the dimensionally reduced codimension-2 action obtained from this
regularized brane is related to the derivatives of the bulk fields
in the near-brane limit. Finally, we show how to define a
renormalized brane action that gives finite results as the size of
the regularizing codimension-1 brane shrinks to zero. We derive RG
equations for this action and show that they agree with those
obtained in special cases by earlier authors using graphical
methods.

Along the way we derive a constraint that directly relates the
on-brane curvature to the brane action, that generalized to higher
codimension the well-known modifications to the Friedmann equation
for codimension-1 branes. However we argue that in the limit of a
very small brane this equation is better understood as a condition
that dynamically determines the size of the regulating
codimension-1 brane as a function of the observable fields in the
problem, rather than as a direct constraint on the on-brane
curvature (since its curvature dependence arises to subleading
order in the low-energy expansion).

For codimension-2 branes the main consequence of this constraint
is instead to directly relate the codimension-2 brane tension,
$\ctwoTensn(\phi)$, to the brane contribution, $\ctwoPotl(\phi)$,
to the effective potential that governs its contribution to the
on-brane curvature. Working perturbatively in the bulk
gravitational coupling, $\kappa^2$, the relation becomes $4 \pi
\ctwoPotl \simeq \kappa^2 \left( {\ctwoTensn}' \right)^2$. A
remarkable consequence of this line or argument is the observation
that any dynamics that allows the bulk scalar field, $\phi$, to
adjust its value at the brane position to minimize its
contribution to the on-brane curvature automatically also
minimizes its coupling to the codimension-2 brane tension (and
vice versa).

We organize our presentation as follows. First, in \S2, we review
the action and field equations for scalar-tensor theory in $D =
d+2$ dimensions. We also summarize the most general solutions to
these equations in the limit that the bulk scalar potential
vanishes, which typically govern the near-brane asymptotics of the
bulk configurations. This allows us to display the singularities
these solutions have as they approach these sources. \S3 then
describes the codimension-1 regularization procedure for dealing
with these singularities, together with the implications of the
Israel junction conditions. \S4 then defines the codimension-2
effective actions for this system, and how they relate to the
asymptotic near-brane behaviour of the bulk fields. Finally, \S5
shows how to renormalize the near-brane divergences. Our
conclusions are summarized in \S6.

\section{The Bulk}

We illustrate the logic of our construction using a simple
higher-dimensional scalar-tensor theory, whose properties we now
briefly describe.

\subsection{Field equations}

Consider therefore the following bulk action, describing the
couplings between the extra-dimensional Einstein-frame metric,
$g_{\ssM\ssN}$, and a real scalar field, $\phi$, in $D = d+2$
spacetime dimensions:\footnote{We use a `mostly plus' signature
metric and Weinberg's curvature conventions \cite{GnC} (that
differ from MTW \cite{MTW} only in the overall sign of the Riemann
tensor).}
\be
    S_B = - \int \d^{D}x \sqrt{-g} \; \left\{ \frac{1}{2\kappa^2}
    \, g^{\ssM\ssN} \Bigl( {\cal R}_{\ssM\ssN} + \partial_\ssM \phi \,
    \partial_\ssN \phi \Bigr) + V(\phi) \right\} \,,
\ee
where ${\cal R}_{\ssM\ssN}$ denotes the Ricci tensor built from
$g_{\ssM\ssN}$. The bulk field equations obtained from this action
are
\ba
    \Box \phi - \kappa^2 \, V'(\phi) &=& 0 \nn\\
    {\cal R}_{\ssM\ssN} + \partial_\ssM \phi \, \partial_\ssN \phi
    + \frac{2\kappa^2 }{d} \, V \, g_{\ssM\ssN} &=& 0 \,.
\ea

Assume, for simplicity, a metric of the form
\ba \label{metricform}
    \d s^2 &=& \d \rho^2 + \hg_{mn} \, \d x^m \, \d x^n \nn\\
    &=& \d \rho^2 + e^{2 B} \, \d \theta^2 + \cg_{\mu \nu} \,
    \d x^\mu \, \d x^\nu \\
    &=& \d \rho^2 + e^{2 B} \, \d \theta^2 + e^{2W} g_{\mu \nu}
    \, \d x^\mu \, \d x^\nu \,,\nn
\ea
where $\theta \simeq \theta + 2\pi$ is an angular coordinate, $B$
and $W$ are functions of $\rho$ only, and $g_{\mu\nu}$ is a
maximally symmetric Minkowski-signature metric depending only on
$x^\mu$. The bulk Ricci tensor then becomes
\ba \label{Riccitensor1}
    {\cal R}_{\mu\nu} &=& \left\{ \frac{\cR}{d} +
    W'' + d \, (W')^2 + W' B' \right\} \, \cg_{\mu\nu} \nn\\
    {\cal R}_{\theta\theta} &=& \Bigl\{ B'' + (B')^2
    + d\, W' B' \Bigr\} \,
    \cg_{\theta\theta} \\
    {\cal R}_{\rho\rho} &=& d\, \Bigl\{ W'' + (W')^2 \Bigr\}
    + B'' + (B')^2 \,.\nn
\ea
so if $\phi = \phi(\rho)$ we obtain the following bulk field
equations:
\ba \label{bulkrhoeqns}
    \phi'' + \Bigl\{ d \, W' + B' \Bigr\} \phi' - \kappa^2 V' &=& 0
    \quad \hbox{($\phi$)}\nn\\
    \frac{\cR}{d} +  W'' + d\, (W')^2 + W' B'
    + \frac{2\kappa^2 V}{d}
    &=& 0 \quad \hbox{($\mu\nu$)} \nn\\
    B'' + (B')^2 + d\, W' B' + \frac{2\kappa^2 V}{d} &=& 0
    \quad \hbox{($\theta\theta$)} \nn\\
    d\, \Bigl\{ W'' + (W')^2 \Bigr\} + B'' + (B')^2 +
    (\phi')^2 + \frac{2\kappa^2 V}{d} &=& 0
    \quad \hbox{($\rho\rho$)} \,.
\ea
In these equations primes indicate differentiation with respect to
the natural argument ({\it i.e.} $\d/\d\phi$ for $V(\phi)$, but
$\d/\d\rho$ for $W(\rho)$, {\it etc.}).

\subsubsection*{The special case $V=0$}

The case $V=0$ is of special interest for several reasons. First,
as we see explicitly below, the field equations may in this case
be explicitly integrated for the axially symmetric ansatz given
above. Second, these $V=0$ solutions often capture the near-brane
behaviour of the bulk fields even when $V$ is nonzero, since in
this limit the potential term is often subdominant to others in
the field equations.

Two classical symmetries of the field equations also emerge when
$V=0$. The first of these is the axion symmetry, for which the
action is unchanged under the replacement
\be \label{axionsymmetry}
    \phi \to \phi + \zeta \,,
\ee
where $\zeta$ is an arbitrary constant and $g_{\ssM\ssN}$ is held
fixed. The second follows from the action's scaling property
$S_\ssB \to \lambda^{d} S_\ssB$ under the replacement
\be \label{scalesymmetry}
    g_{\ssM\ssN} \to \lambda^2 g_{\ssM\ssN} \,,
\ee
with constant $\lambda$ and $\phi$ held fixed. Both of these
symmetries take solutions of the classical field equations into
distinct new solutions of the same equations.

\subsection{Axisymmetric bulk solutions}

The bulk field equation can be integrated to obtain the general
solution in the special case $V = 0$, and we collect these
solutions in this section. As discussed above, these solutions are
also relevant when $V \ne 0$, since even in this case they can
capture the asymptotic behavior of bulk solutions very near the
branes which source them.

When $V=0$ (or when $V$ is minimized at $V=0$) a trivial solution
is $\phi' = W' = 0$ and $g_{\mu\nu} = \eta_{\mu\nu}$, but $e^B =
\alpha \rho$. The constant $\alpha$ can be absorbed by re-scaling
it into the coordinate $\theta$, but only at the expense of
changing its periodicity to $\theta \simeq \theta + 2\pi\alpha$,
showing that this solution corresponds to flat space (in
cylindrical coordinates) when $\alpha = 1$, or a cone (with
conical singularity at $\rho = 0$ and defect angle $2\pi \delta$,
with $\delta = 1 - \alpha$) if $\alpha \ne 1$.

The general solution to the dilaton and Einstein equations (see
Appendix \ref{App:SolvingFieldEq} for details) when $V=0$ is
\be \label{generalsoln}
    e^\phi = e^{\phi_0} \left(\frac{r}{r_0} \right)^{p_\phi}
    \,, \quad
    e^B = e^{B_0} \left(\frac{r}{r_0} \right)^{p_\ssB} \,,
\ee
and
\be \label{generalsolnW}
    e^{(d-1)W}
    = \frac{(r_0/l_\ssW)^\Omega
    + (l_\ssW/r_0)^\Omega}{(r/l_\ssW)^\Omega
    + (l_\ssW/r)^\Omega}
    \left(\frac{r_0}{r}\right)^{p_\ssB} \, e^{(d-1) W_0}\,,
\ee
where
\be \label{omegaconstraint}
 \Omega^2 = p_\ssB^2 + p_\phi^2\left( \frac{d-1}{d}
 \right)  \,,
\ee
and we may take the positive root without loss of generality. The
freedom to re-scale $x^\mu$ allows us to shift $W_0$ arbitrarily,
and re-scalings of $r$ allow any value to be chosen for $r_0$,
leaving five quantities $l_\ssW/r_0$, $\phi_0$, $B_0$, $p_\phi$
and $p_\ssB$ as the remaining integration constants. The radial
coordinate, $r$, used to solve the equations is related to the
radial proper distance, $\rho$, by
\be \label{rdef}
    \frac{\d r}{r} = \xi \, e^{-B - d\,W} \, \d \rho \,,
\ee
for arbitrary constant $\xi$.

The curvature scalar in the brane directions is given in terms of
the above constants by
\be
    R = \frac{4 d \xi^2 \Omega^2}{(d-1) \, r_0^2 }
    \;
    \left[ \left( \frac{r_0}{l_\ssW} \right)^\Omega +
    \left( \frac{l_\ssW}{r_0} \right)^\Omega
    \right]^{-2} e^{-2(d-1)W_0}  \ge 0\,.
\ee
Notice that the curvature obtained is strictly non-negative
(corresponding in our conventions to flat or anti-de Sitter
geometries), in agreement with general no-go arguments for finding
de Sitter solutions in higher-dimensional supergravity.

A key feature of these solutions is the singularities they
generically display as $r \to 0$ and $r \to \infty$, which we
interpret as being due to the presence there of source branes
having dimension $d = D-2$. This divergent near-brane behaviour is
an important departure from the codimension-one case. Furthermore,
even though these asymptotic near-brane forms are derived using
$V=0$, the singular behaviour given above often provides a good
approximation in the near-brane limit even for nonzero $V$. To see
why, consider the example of a potential of the form $V(\phi) =
V_0 \, e^{\lambda \phi}$. Evaluated at the solution of
eq.~\pref{powerlaws}, this gives the following contributions to
the field equations
\be
    \kappa^2 V' \propto \lambda \kappa^2 V \propto
    \lambda r^{\lambda p_\phi} \propto \lambda \rho^\zeta \,,
\ee
for calculable $\zeta$. The main point is that this often
represents a subdominant contribution to equations
eqs.~\pref{bulkrhoeqns} as $\rho \to 0$ near the brane, provided
$\zeta > -2$, since the other terms in these equations vary like
$\partial_\rho^2\phi \propto 1/\rho^2$.

\subsubsection*{Special Cases}

There are a number of special cases that are of particular
interest in what follows.

\medskip\noindent{\em Conical Singularity:}

\medskip\noindent
If we wish to avoid a curvature singularity at $r=0$ we must take
$p_\phi = 0$ and so $p_\ssB = \Omega := p$, in which case $\phi =
\phi_0$ and $e^B = e^{B_0}(r/r_0)^{p}$. The warp factor then
becomes
\be \label{nonsing1}
    e^{(d-1)W} =  \left( \frac{r_0^{2p} + l_\ssW^{2p}}{r^{2p}
    + l_\ssW^{2p}} \right) e^{(d-1) W_0}\,,
\ee
where $r_0$ is an arbitrary point where the metric functions are
assumed known: $B(r_0) = B_0$ and $W(r_0) = W_0$. The proper
distance, $\rho$, is then related to $r$ by
\be \label{pdtake2}
    \xi \, \exd \rho = e^{B_0+dW_0} \left( \frac{r}{r_0} \right)^p
    \left( \frac{r_0^{2p} + l_\ssW^{2p}}{r^{2p} + l_\ssW^{2p}}
    \right)^{d/(d-1)} \frac{\exd r}{r}   \,,
\ee
which shows that $p\, \xi \rho = e^{B_0 +dW_0} \left[1 + (r_0 /
l_\ssW)^{2p} \right]^{d/(d-1)} (r/r_0)^p + {\cal O}\left( r^{3p}
\right)$ near $r = 0$, and so $e^B = \alpha \rho + {\cal
O}(\rho^3)$ with $\alpha = p\, \xi e^{-dW_0} \left[ 1 + (r_0 /
l_\ssW)^{2p} \right]^{-d/(d-1)}$.

The curvature similarly reduces to
\ba \label{nonsing2}
  R &=& \frac{4 d \, p^2 \xi^2}{(d-1) \, r_0^2} \,
  \left( \frac{r_0}{l_\ssW} \right)^{2p} \left[
  1 + \left( \frac{r_0}{l_\ssW} \right)^{2p} \right]^{-2}
  e^{-2(d-1)W_0} \nn\\
  &=& \frac{4 d \, \alpha^2}{(d-1) \, r_0^2} \,
  \left( \frac{r_0}{l_\ssW} \right)^{2p} \left[
  1 + \left( \frac{r_0}{l_\ssW} \right)^{2p} \right]^{2/(d-1)}
  e^{2W_0}\,.
\ea
In general, this geometry has a conical singularity at $\rho=0$,
whose defect angle is $2\pi \delta = 2\pi(1 - \alpha)$. When
$\alpha = 1$ it is instead purely a coordinate singularity, which
requires $p\, \xi = e^{dW_0} \left[ 1 + \left( r_0/ l_\ssW
\right)^{2p} \right]^{d/(d-1)}$.

\medskip\noindent{\em Flat Brane Geometries:}

\medskip\noindent
As shown in more detail in Appendix \ref{App:SolvingFieldEq} when
the induced brane geometry is flat ($R=0$), the solutions have a
simple form when written in terms of $\rho$:
\be \label{powerlaws}
    e^\phi = e^{\phi_0} \left( \frac{\rho}{\rho_0}
    \right)^{\gamma} \,, \quad
    e^B = \alpha \, \rho_0 \left( \frac{\rho}{\rho_0}
    \right)^{\beta}
    \quad \hbox{and}\quad
    e^{W} = e^{W_0} \left( \frac{\rho}{\rho_0}
    \right)^{\omega} \,,
\ee
where the powers satisfy
\be \label{kasnerconditions}
    d \, \omega^2 + \beta^2 + \gamma^2
    = d \, \omega + \beta = 1\,.
\ee
In terms of these constants, the trivial solution given above
corresponds to the choices $\omega = \gamma = 0$ and $\beta = 1$.
For more general powers the bulk geometry potentially has
singularities at $\rho = 0$ and at $\rho \to \infty$, which we
interpret as being due to the presence there of codimension-2
branes. Several special subcases are worth identifying.
\begin{itemize}
\item {\it Conical singularity:} The singularity of the bulk
geometry at $\rho = 0$ is a conical singularity (as opposed to a
curvature singularity), if and only if $\beta = 1$. In this case
eqs.~\pref{kasnerconditions} imply $\omega = \gamma = 0$, implying
$\phi$ and $W$ are constant and $e^B = \alpha \, \rho$. This
corresponds to the limit $l_\ssW \to \infty$ of the previous
example, and as before the conical defect angle, $2\pi \delta$,
satisfies $\delta = 1- \alpha$.
\item {\it Constant dilaton:} The scalar $\phi$ does not vary
across the extra dimensions if and only if $\gamma = 0$, in which
case eqs.~\pref{kasnerconditions} admit two solutions for $\omega$
and $\beta$: ($i$) the conical solution just discussed, $\omega =
0$ and $\beta = 1$; or ($ii$) the curved geometry with $\omega =
2/(d+1)$ and $\beta = -(d-1)/(d+1)$. Notice that negative $\beta$
implies the circumferences of circles in the extra dimensions
having radius $\rho$ {\em decreases} with increasing $\rho$ rather
than increasing.
\end{itemize}

\EPSFIGURE[t]{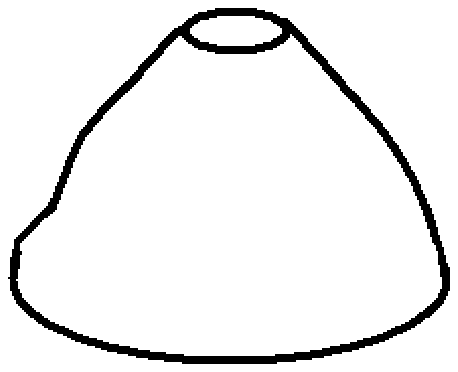, width = 0.4\textwidth,
height=5cm}{\sl A cartoon of the near-brane cap
geometry.\label{fig:regcap}}

\pagebreak
\section{The Codimension-One Crutch}

We turn now to the problem of establishing how the asymptotic
features of the singular near-brane bulk fields are related to the
properties of the effective codimension-2 brane which does the
sourcing. Experience with the related problem of finding the
electrostatic field sourced by a localized charge distribution, we
expect to find the near-brane power-law behaviour of the bulk
field to be related to the physical properties of the brane.

A trick for finding this connection between a small source brane
and the bulk field to which it gives rise involves resolving the
codimension-2 singularity in terms of a codimension-1 brane having
a very small proper circumference \cite{UVCaps,BdRHT}. For
instance for the singularity near $\rho = 0$, we replace the
geometry for $\rho < \rho_b$ with a new smooth geometry (see
Fig.~\ref{fig:regcap}). The boundary between these two geometries
represents the codimension-1 brane, whose properties can be
related to the inner and outer geometries using standard junction
conditions.\footnote{For completeness the derivation of these
conditions is summarized in our conventions in Appendix
\ref{App:jumpconds}.} In making this model we expect to derive
connections between the bulk and codimension-2 brane that are more
robust than the details of this particular codimension-1
realization.

The rest of this section collects the results of such a
junction-condition analysis. The first step is to more
specifically identify the exterior (`bulk') and interior (`cap')
geometries, and then to choose a codimension-1 brane action whose
structure is sufficiently rich to allow independent contributions
to the two independent stress-energy components, $T_{\mu\nu}$ and
$T_{\theta\theta}$, that the matching between the two geometries
requires. How these two stress-energies show up physically in the
low-energy codimension-2 brane effective action is then identified
in a subsequent section, \S4.

\subsection{Bulk Properties}
\label{extint}

We start with a discussion of the relevant geometries.

\subsubsection*{Interior geometry}

Inside the circular brane we assume a nonsingular configuration
that matches properly to the exterior solution. When $V = 0$ this
solution may be obtained explicitly from the conical solution
described in the previous section, with $x^\mu$ re-scaled to
ensure $W(0) = 0$. That is, we take $W_0 = r_0 = p_\phi = 0$ and
$\xi = p_\ssB = 1$, and so
\be \label{interiorgeometry}
    \phi_i = \phi_b \,, \quad
    e^{B_i} =  r
    \qquad \hbox{and} \qquad
    e^{(d-1)W_i} = \frac{l_{\ssW i}^2}{r^2 + l_{\ssW i}^2} \qquad
    \hbox{for $0 < r < r_{b}$} \,,
\ee
for constants $\phi_b$, $l_{\ssW i}$ and $r_b$. The coordinate $r$
is connected to proper distance, $\rho$, by the relation
\be \label{relrhorint2}
 \exd \ln{r} = e^{-B_i-d W_i} \, \exd \rho
 \hskip0.5cm \hbox{and so} \hskip0.5cm
 \rho = \int_0^{r} \exd \hat r \; e^{dW_i(\hat r)} \,.
\ee
At the codimension-1 brane position we have $\phi = \phi_b$,
$e^{B(r_b)} = r_b$ and $e^{(d-1)W_i(r_b)} = l_{\ssW i}^2/(r_b^2 +
l_{\ssW i}^2)$. The derivatives relevant to the junction
conditions (more about which later) are
\be
    r \partial_{r} \phi_i = 0 \,, \quad
    r \partial_{r} B_i = 1 \quad \hbox{and} \quad
    r \partial_{r}  W_i = - \left( \frac{2}{d-1} \right)
    \frac{r^2}{r^2 + l_{\ssW i}^2} \,.
\ee

Finally, the scalar curvature of the $d$ directions parallel to
the brane is related to the constant $l_{\ssW i}$ by
\be
    R = \frac{4 d}{(d-1)\, l_{\ssW i}^2} \,,
\ee
so we can trade the integration constant $l_{\ssW i}$ for the
on-brane spatial curvature, $R \ge 0$. Notice that since $l_{\ssW
i}$ is of order the radius of curvature of $R$, while $r_b$ is
microscopic, our interest is in the regime $r_b \ll l_{\ssW i}$.
In this limit the warp factor never strays far from unity,
$e^{-(d-1) W_i(r_b)} = 1 + r_b^2/l_{\ssW i}^2$, and so
\be
 \rho_b = \int_0^{r_b} \exd r \; e^{dW_i} = r_b
 \left[ 1 - \frac{d}{3(d-1)} \left( \frac{r_b}{l_{\ssW i}}
 \right)^2 + \cdots \right]
 = r_b \left[ 1 - \frac{1}{12} \, r_b^2 R + \cdots \right]\,.
\ee

\subsubsection*{Exterior geometry}

Outside the brane we take the exterior configuration to be a
general geometry described by functions $\phi_e$, $W_e$ and $B_e$,
which we only assume solves the bulk field equations. In
particular these equations could include the bulk potential $V$.
Although much of what follows does not require knowing the
explicit form of the solution in detail, for concreteness' sake it
is also worth keeping some explicit external solutions in mind.
When this is useful we use the $V=0$ solutions described above,
assuming the contribution of $V$ can be ignored very close to the
brane.

To describe the exterior solutions we extend both the proper
distance, $\rho$, and coordinate $r$ outside the brane. However,
unlike for the interior solutions, for the exterior solutions the
requirement that the brane position be located at $r=r_b$ removes
the freedom to place the potential singularity at $r = 0$. In this
case, repeating the arguments of appendix
\ref{App:SolvingFieldEq}, suggests defining $r$ in the exterior
region by the relation
\be
 \xi \, \exd \rho =
 e^{B_e + d W_e} \, \exd \ln (r-l)\,,
\ee
where the choice $\xi = r_b/(r_b - l)$ ensures $\exd \rho/\exd r$
remains continuous across $r = r_b$.

This leads to the solutions
\be \label{generalsolnshft}
    e^{\phi_e} = e^{\phi_b} \left(\frac{r - l}{r_b - l}
    \right)^{ p_\phi}
    \,, \quad
    e^{B_e} = r_b \left(\frac{r - l}{r_b - l}
    \right)^{p_\ssB} \,,
\ee
and
\be \label{generalsolnWshft}
    e^{(d-1)W_e} = \frac{[(r_b - l)/l_\ssW]^\Omega
    + [l_\ssW/(r_b - l)]^\Omega}{[(r - l)/l_\ssW]^\Omega
    + [l_\ssW/(r - l)]^\Omega}
    \left(\frac{r_b -l}{r -l}\right)^{p_\ssB}
    \frac{l_{\ssW i}^2}{r_b^2
    + l_{\ssW i}^2}\,,
\ee
with $\Omega$ given by eq.~\pref{omegaconstraint} as before. In
writing these we use three of the integration constants to ensure
that these functions are continuous with the interior solution
across the brane at $r=r_b$. All expressions are nonsingular
provided $r_b > l$ (where $l$ can be negative) because they apply
only for $r > r_b$.

Of particular interest in what follows is the regime where $r$,
$l_{\ssW i}$ and $l_\ssW$ are all much greater than $r_b$ and
$|l|$, in which case --- keeping in mind $\Omega \ge 0$, and
$\Omega = 0$ if and only if $p_\ssB = p_\phi = 0$ --- the
expression for $W_e$ simplifies to
\be \label{generalsolnWshftapp}
    e^{(d-1)W_e} \simeq \frac{(l_\ssW/
    r_b)^\Omega}{(r/l_\ssW)^\Omega
    + (l_\ssW/r)^\Omega}
    \left(\frac{r_b}{r}\right)^{p_{\ssB}}
    \,.
\ee

A final continuity condition comes from the requirement that the
external geometry reproduce the value for $R$ given by the cap,
which requires $l_\ssW$ (say) to be chosen to satisfy
\ba \label{metricmatching}
 \frac{(d-1) R}{4\, d} = \frac{1}{l_{\ssW i}^2}
 &=& \frac{\xi^2 \Omega^2/(r_b-l)^2 }{
 \left\{ \left[ {(r_b-l)}/{l_\ssW}
 \right]^\Omega + \left[ {l_\ssW}/{(r_b-l)} \right]^\Omega
 \right\}^{2}}   \left( \frac{r_b^2
    + l_{\ssW i}^2}{l_{\ssW i}^2} \right)^2 \nn\\
 &\simeq& \frac{\xi^2 \Omega^2}{(r_b-l)^2}
 \left( \frac{r_b-l}{l_\ssW} \right)^{2\Omega} \,,
\ea
and so $(r_b-l)/l_{\ssW i} \simeq \xi \Omega [(r_b -l) /l_\ssW
]^{\Omega}$. Here the final approximate equality assumes $r_b \ll
l_{\ssW i}$ and $r_b - l \ll l_\ssW$.

For later purposes, the relevant derivatives are
\be \label{genderivshft}
    \frac{1}{\xi} \, e^{B_e + d W_e} \partial_\rho \phi_e
    = \frac{\partial\,\phi_e}{\partial \ln (r-l)}
    = p_\phi  \,, \quad
 \frac{1}{\xi} \, e^{B_e + d W_e} \partial_\rho B_e =
 \frac{\partial\,B_e}{\partial \ln{(r-l)}}
     = p_B  \,,
\ee

and
\be \label{genderivshftW}
    \frac{(d-1)}{\xi} \, e^{B_e + d W_e} \partial_\rho W_e
    = (d-1)  \frac{\partial\,W_e}{
    \partial \ln (r-l)} = - \left\{ p_\ssB + \,\Omega
    \left[ \frac{(r-l)^{2\Omega}
    - l_\ssW^{2\Omega}}{(r-l)^{2\Omega}
    + l_\ssW^{2\Omega}} \right] \right\}
     \,.
\ee
These derivatives are not continuous when matched to the interior
solutions at $r=r_b$, and the resulting discontinuity is related
by the junction conditions to the properties of the codimension-1
brane located at this position.

\medskip\noindent{\it Flat Branes}

\smallskip\noindent Of special importance is the special case
of flat induced brane geometries, $R=0$, as obtained by taking
$l_{\ssW i} \to \infty$, since these include many of the
best-studied examples. In this case the warping in the cap becomes
a constant, $W_i = 0$, and because the metric matching condition,
eq.~\pref{metricmatching}, also implies $l_\ssW \to \infty$, the
exterior solutions reduce to
\be \label{powerlawsshftr}
    e^{\phi_e} = e^{\phi_b} \left( \frac{r-l}{r_b-l}
    \right)^{p_\phi} \,, \quad
    e^{B_e} = r_b\, \left( \frac{r-l}{r_b-l}
    \right)^{p_\ssB}
    \quad \hbox{and}\quad
    e^{(d-1)W_e} = \left( \frac{r-l}{r_b-l}
    \right)^{\Omega - p_\ssB} \,.
\ee

Keeping in mind $\xi = r_b/(r_b - l)$, the proper distance in this
case satisfies
\be
 \exd \rho = (r_b - l) \left( \frac{r-l}{r_b-l}
 \right)^{(-p_\ssB + d\Omega)/(d-1)} \exd \ln (r-l) \,,
\ee
and so $\rho-\ell \propto (r-l)^{(-p_\ssB + d\Omega)/(d-1)}$,
where the integration constant $\ell$ is defined so that $\rho$
would approach $\ell$ in the limit $r \to l$ if their relation
were defined by the exterior solution for all $r$. Notice that
$\ell$ can be negative. In terms of $\rho$ the solutions become
\be \label{powerlawsshft}
    e^{\phi_e} = e^{\phi_b} \left( \frac{\rho - \ell}{\rho_b
    - \ell} \right)^{\gamma} \,, \quad
    e^{B_e} = \rho_b\, \left( \frac{\rho - \ell}{\rho_b - \ell}
    \right)^{\beta}
    \quad \hbox{and}\quad
    e^{W_e} = \left( \frac{\rho - \ell}{\rho_b - \ell}
    \right)^{\omega} \,,
\ee
where we use $r_b = \rho_b$ when $R=0$ in evaluating
$B_e(\rho_b)$, and as before the powers $\gamma$, $\beta$ and
$\omega$ satisfy eqs.~\pref{kasnerconditions}.

In the even more special case where $\omega = \gamma = 0$ and
$\beta = 1$, we have $p_\phi = 0$ and $p_\ssB = \Omega = 1$, and
so $\exd \rho = \exd (r - l)$, implying $\ell = l$. In this case
the exterior solution becomes a conical space, whose metric can be
written as
\begin{eqnarray}
 \exd s^2_{e} &=& \eta_{\mu\nu} \, \exd x^{\mu} \exd x^{\nu}
 + \exd \rho^2 + e^{2 B_e} \, \exd \theta^2 \nonumber \\
 &=& \eta_{\mu\nu} \, \exd x^{\mu} \exd x^{\nu}
 + \exd \varrho^2 + \alpha^2 \, \varrho^2
 \exd \theta^2 \,,\label{limconsp}
\end{eqnarray}
where $\varrho = \rho - \ell$, revealing the defect angle $2\pi
\delta = 2\pi (1-\alpha)$, with
\be\label{alfadef}
 \alpha = \frac{\rho_b}{\rho_b - \ell} > 0
 \quad \hbox{and so} \quad
 \delta = - \frac{\ell}{\rho_b - \ell} \,.
\ee
Evidently $\alpha < 1$ and $\delta > 0$ if $\ell < 0$ while
$\alpha > 1$ and $\delta < 0$ if $\ell > 0$. Notice that because
$\rho_b = r_b$ when $R = 0$ it follows that $\alpha=\xi$, in
agreement with the discussion of the $R=0$ conical solution just
below eq.~\pref{pdtake2}. Since $\alpha$ is a simply measured
parameter characterizing the exterior geometry, it is convenient
to regard the above relation as defining the quantity $l$ (or
$\ell$) in terms of $r_b$ and $\alpha$.

\subsubsection*{Extrinsic curvatures}

The extrinsic curvature of the surfaces of constant $\rho$ in the
metric of eq.~\pref{metricform} is $K_{mn} = \frac12 \,
\partial_\rho \hg_{mn}$, whose components are
\ba
    K_{\mu\nu} &=& W' \cg_{\mu\nu} = W' e^{2W} g_{\mu\nu} \nn\\
    K_{\theta\theta} &=& B' g_{\theta\theta} = B' e^{2B} \,,
\ea
and whose trace, $K = \hg^{mn} K_{mn} = \cg^{\mu\nu} K_{\mu\nu} +
g^{\theta\theta} K_{\theta\theta}$, is
\be
    K = d \, W' + B' \,.
\ee
As before, primes denote derivatives with respect to $\rho$,
and we reserve overdots to denote differentiation with respect to
$r$: $\phi' := \partial_\rho \phi$ and $\dot \phi := \partial_r
\phi$.

The Gauss-Codazzi equations give the $D = (d+2)$--dimensional
Riemann tensor in terms of the $(d+1)$ dimensional Riemann tensor
and the extrinsic curvature, which for the metric
\pref{metricform} becomes
\ba \label{Riccitensor2}
    {\cal R}_{\mu\nu} &=& \cR_{\mu\nu} + \partial_\rho K_{\mu\nu} - 2 K_{\mu\lambda}
    {K^\lambda}_\nu + K \, K_{\mu\nu} \nn\\
    &=& \cR_{\mu\nu} + \Bigl\{ W'' + d\, (W')^2 + W' B' \Bigr\}
    \, \cg_{\mu\nu} \nn\\
    {\cal R}_{\theta\theta} &=&  \partial_\rho K_{\theta\theta}
    - 2 g^{\theta\theta} (K_{\theta\theta})^2 + K \, K_{\theta\theta} \nn\\
    &=& \Bigl\{ B'' + (B')^2 + d\, W' B' \Bigr\} \,
    \cg_{\theta\theta} \nn\\
    {\cal R}_{\rho\rho} &=& \partial_\rho K + K_{\mu\nu} K^{\mu\nu}
    + K_{\theta\theta} K^{\theta\theta} \nn\\
    &=& d\, \Bigl[ W'' + (W')^2 \Bigr] + B'' + (B')^2 \,,
\ea
in agreement with eqs.~\pref{Riccitensor1}.

\subsection{Junction Conditions}

The equations of motion at the codimension-1 brane consist of the
requirements of continuity for $g_{\ssM\ssN}$ and $\phi$, as well
as a set of `jump' conditions relating the functional derivatives
of the brane action, $S_b$, with discontinuities in the radial
derivatives of the bulk fields (see Appendix \ref{App:jumpconds}
for details).

\subsubsection*{Metric jump conditions}

In terms of the brane stress energy,
\be
    t^{mn} \equiv \frac{2}{\sqrt{-\hg}} \; \frac{\delta S_b}{\delta
    \hg_{mn}} \,,
\ee
the metric discontinuity condition is given by the Israel junction
condition
\be \label{IsraelJCs}
    \Bigl[ K_{mn} - K \, \hg_{mn} \Bigr]_b + \kappa^2 \,
    t_{mn} = 0
    \,,
\ee
where we define $[A]_b := A(\rho_b + \epsilon) - A(\rho_b -
\epsilon)$, with $\epsilon \to 0$. Using the metric of
eq.~\pref{metricform}, this leads to
\ba
    \Bigl[ (d-1) W' + B' \Bigr]_b \cg_{\mu\nu}
    &=& \kappa^2 t_{\mu\nu}  \nn\\
    \Bigl[ d\, W' \Bigr]_b g_{\theta\theta}
    &=& \kappa^2 t_{\theta\theta} \,,
\ea
which in particular implies
\be
    \Bigl[ W' - B' \Bigr]_b = \kappa^2 \left( g^{\theta\theta}
    t_{\theta\theta} - \frac{1}{d} \, \cg^{\mu\nu} t_{\mu\nu}
    \right) \,.
\ee
This last equation shows that $\Bigl[ W' - B' \Bigr]_b = 0$ across
a brane for which the codimension-1 stress energy is pure tension:
$t_{\mu\nu} = T \cg_{\mu\nu}$ and $t_{\theta\theta} = T
g_{\theta\theta}$.

\subsubsection*{Scalar jump condition}

For the scalar field the corresponding jump condition relates the
$\phi$-dependence of the brane action to the jump of $\phi'$
across the brane. That is
\be \label{scalarjump}
    \Bigl[ \phi' \Bigr]_b + \frac{\kappa^2}{\sqrt{-\hg}} \,
    \frac{\delta S_b}{\delta \phi} = 0 \,.
\ee
In what follows it proves to be of interest to consider variations
for which the induced metric at the brane position varies as
$\phi$ does. Eq.~\pref{scalarjump} also applies in this case,
provided the $\phi$-variation of the induced metrics that are
implicit in $S_b$ are also included when computing the variational
derivative on its right-hand side (see Appendix
\ref{App:jumpconds}).

\subsection{The codimension-1 brane action}

To make the discussion explicit we use a codimension-1 brane
action which includes a massless brane scalar degree of freedom,
$\sigma$, that couples to the bulk fields through the
action\footnote{See appendix \ref{App:MatchingwDerivatives} for a
discussion of matching with higher-derivative terms in the brane
action.}
\be \label{Cod1Sb}
    S_1 = - \int \d^{d+1} x \sqrt{-\hg} \; \left\{ \coneTensn (\phi)
    + \frac12 \, \coneKin (\phi) \, \hg^{mn} \partial_m \sigma
    \partial_n \sigma \right\} \,.
\ee
Notice that this brane action generically breaks both of the
symmetries (discussed in \S2 above) that the bulk equations
acquire when $V=0$. In particular, eq.~\pref{axionsymmetry} is
broken if and only if either $\coneTensn $ or $\coneKin $ depends
on $\phi$, while the scaling symmetry, eq.~\pref{scalesymmetry},
is broken by any (even a constant) nonzero $\coneTensn $ or
$\coneKin $. A `diagonal' combination of these two does survive
the inclusion of the brane action in the special case $\coneTensn
(\phi) = A \, (\phi) = A \, e^{a\phi}$ and $\coneKin (\phi) = B \,
e^{b\phi}$, since these choices preserve the combination
$g_{\ssM\ssN} \to \lambda^2 g_{\ssM\ssN}$ and $\phi \to \phi +
\zeta$ provided $b = -a$ and $e^{b\zeta} = \lambda$.

We use the brane scalar field, $\sigma$, as a trick to generate an
independent stress energy, $t_{\theta\theta}$, in the $\theta$
direction, in order to distinguish its low-energy implications
from those of the on-brane stress energy, $t_{\mu\nu}$. This can
be done if $\sigma$ takes values on a circle, $\sigma \simeq
\sigma + 2\pi$, since we can solve the $\sigma$ equation of motion
\be
    \hat\nabla_m \Bigl[ \coneKin (\phi) \, \hg^{mn} \hat \nabla_n
    \sigma \Bigr] = 0 \,,
\ee
in a sector where it winds nontrivially around the brane:
\be
    \sigma = n \,\theta \,,
\ee
where $n$ is an integer.

The stress energy produced by this action is
\be
    t^{mn} \equiv \frac{2}{\sqrt{-\hg}} \; \frac{\delta S_b}{\delta
    \hg_{mn}} = - \hg^{mn} \left\{ \, \coneTensn (\phi) + \frac12 \, \coneKin (\phi)
    \, \partial_p\sigma \partial^p\sigma \right\} + \coneKin (\phi) \,
    \partial^m\sigma
    \partial^n\sigma \,,
\ee
which, when evaluated with $\partial_\theta \sigma = n$ leads to
\ba
    t_{\mu\nu} &=& - \left\{ \coneTensn  + \frac{
    n^2}{2} \, e^{-2B} \, \coneKin   \right\} \, \cg_{\mu\nu}
    = - \left\{ \coneTensn  + \frac{
    n^2}{2 r_b^2} \, \coneKin   \right\} \, \cg_{\mu\nu} \nn\\
    t_{\theta\theta} &=& - \left\{ \coneTensn  - \frac{
    n^2}{2} \, e^{-2B} \, \coneKin  \right\} \, g_{\theta\theta}
    = - \left\{ \coneTensn  - \frac{
    n^2}{2 r_b^2 } \, \coneKin  \right\} \, g_{\theta\theta}\,.
\ea
In each case the second equality uses continuity of the metric at
the brane position,
\be
 e^{-(d-1) W(r_b)} = 1 + \frac{d-1}{4d} \, r_b^2 R
 \quad \hbox{and} \quad
 e^{B(r_b)} = r_b = \rho_b \left[ 1 + \frac{1}{12} \,
 \rho_b^2 R + \cdots \right] \,,
\ee
showing that we can replace $r_b$ by $\rho_b$ provided we neglect
subdominant ${\cal O}(r_b^2 R)$ terms.

In what follows an important role is played by the dimensional
reduction of these two quantities on the small circle at $r =
r_b$, defined by:
\ba \label{T2defs}
    \ctwoTensn  = 2\pi \, e^{B_b+dW_b} \left\{ \coneTensn
    + \frac{n^2}{2} \, e^{-2B_b} \, \coneKin   \right\}
    &=& 2\pi r_b \, \left[1 + \frac{d-1}{4d} \, r_b^2 R
    \right]^{-d/(d-1)} \left\{ \coneTensn
    + \frac{n^2}{2r_b^2} \, \coneKin   \right\} \nn\\
    &\simeq& 2\pi \rho_b \left\{ \coneTensn  + \frac{
    n^2}{2 \rho_b^2} \, \coneKin   \right\} \,,
\ea
and
\ba \label{U2defs}
    \ctwoPotl  = -2\pi \, e^{B_b+dW_b} \left\{ \coneTensn
    - \frac{n^2}{2} \, e^{-2B_b} \, \coneKin  \right\}
    &=& -2\pi r_b \, \left[1 + \frac{d-1}{4d} \, r_b^2 R
    \right]^{-d/(d-1)} \left\{ \coneTensn
    - \frac{n^2}{2r_b^2} \, \coneKin   \right\} \nn\\
    &\simeq& -2\pi \rho_b \left\{ \coneTensn  - \frac{
    n^2}{2 \rho_b^2 } \, \coneKin  \right\} \,,
\ea
with the approximate inequalities again using $e^B \simeq \rho_b$
and $W \simeq 0$. Perhaps not surprisingly, $\ctwoTensn $ will
turn out to play the role of the leading approximation to the
effective codimension-2 brane tension that is appropriate to bulk
physics on scales that are large compared to $\rho_b$, the size of
the codimension one crutch. As is shown below, $\ctwoPotl $ has a
similarly clean physical interpretation, being (when $V=0$) the
leading approximation to the brane contribution to the low-energy
potential governing the physics below the KK scale, after all of
the bulk physics has been integrated out.

\subsubsection*{Matching conditions}

We next specialize the general matching conditions to the assumed
axisymmetric bulk geometries and the above codimension-1 brane
action (for details see Appendix \ref{App:SolvingFieldEq}). Using
the metric ansatz, eq.~\pref{metricform}, we wish to track how the
functions $B$, $W$ and $\phi$ change as we cross the brane
position. In what follows we use the explicit form of the
nonsingular interior geometry, eq.~\pref{interiorgeometry}:
$\phi_i = \phi_b$, $e^{(d-1)W_i} = l_{\ssW i}^2/(r^2 + l_{\ssW
i}^2)$ and $e^{B_i} = r$, but for most purposes do not require the
details of the corresponding external solution,
eqs.~\pref{generalsolnshft}.

The three exterior functions are subject to 6 conditions at $r =
r_b$. Three of these conditions express the continuity of $\phi$,
$W$ and $B$,
\be
    \phi_e(r_b) = \phi_b \,, \qquad
    e^{-(d-1)W_e(r_b)} = 1 + \frac{1}{4d} \,
    r_b^2 R \qquad\hbox{and}\qquad
    e^{B_e(r_b)} = r_b \,,
\ee
and have already been used in the explicit expressions for the
exterior solutions in eqs.~\pref{generalsolnshft}. Continuity also
demands the induced brane metric, $g_{\mu\nu}$, must also agree on
both sides of the brane, as must therefore its curvature scalar,
$R$.

There are three independent jump conditions for the geometries of
interest, one each for $\partial_\rho\phi$, $\partial_\rho W$ and
$\partial_\rho B$. Evaluating these with the geometry of interest
leads to the following relations
\ba \label{dilatonJC}
    \Bigl[ \partial_\rho \phi \Bigr]_b &\simeq&
    \frac{\kappa^2}{\rho_b} \left\{ \rho_b \coneTensn (\phi_b)
    + \frac{n^2}{2\rho_b} \, \coneKin (\phi_b) \right\}' \\
    \label{IsraelJCmunu}
    \Bigl[ (d-1) \partial_\rho W + \partial_\rho B \Bigr]_b
    &\simeq& - \kappa^2 \, \left\{ \coneTensn  + \frac{
    n^2}{2 \rho_b^2 } \, \coneKin  \right\} \\
    \label{IsraelJCthetatheta}
    \Bigl[ d\, \partial_\rho W \Bigr]_b &\simeq&
    - \kappa^2 \, \left\{ \coneTensn  - \frac{
    n^2}{2 \rho_b^2 } \, \coneKin  \right\} \,,
\ea
where the approximate equality indicates neglect of powers of
$r_b^2 R$, and the prime in the first line denotes differentiation
with respect to $\phi_b = \phi(\phi_b)$. This derivative is not
taken inside the parenthesis to allow for the possibility that
quantities like $\rho_b$ might acquire a dependence on $\phi_b$
through the solving of the junction conditions.

The explicit exterior solutions, \pref{generalsolnshft}, nominally
depend on six independent parameters: $\phi_b$, $p_\phi$,
$p_\ssB$, $l$, $l_\ssW$ and $r_b$, in terms of which all of the
remaining parameters (like $\ell$, $\rho_b$, $R$, {\it etc.}) can
be expressed. These six are subject to the three junction
conditions, eqs.~\pref{dilatonJC} through
\pref{IsraelJCthetatheta}. Assuming the functions
$\coneTensn(\phi_b)$ and $\coneKin(\phi_b)$ are specified, we
therefore generically expect a three-parameter family of
solutions, corresponding to the freedom to choose the radius,
$\rho_b$, where we place the codimension-1 brane; the on-brane
curvature scalar, $R$; as well as the quantity $\alpha = r_b/(r_b
- l)$.

\subsubsection*{The Brane at Infinity}

A similar story also applies as $r \to \infty$, whose singularity
can also be replaced by an appropriate codimension-1 brane and
cap. The resulting brane therefore turns out to have properties
that are predictable, once one specifies the functions
$\coneTensn$ and $\coneKin$ that define the properties of the
codimension-1 brane; its precise position; as well as the induced
brane curvature scalar, $R$ and $\alpha$ that characterize the
bulk geometry \cite{UVCaps,BdRHT}. That this is true may be seen
from the above connection between brane properties and derivatives
of the bulk fields at the brane positions, together with our
ability to integrate the bulk field equations in the $r$
direction. These imply that the brane at $r=0$ provides a set of
`initial' conditions at $r=r_b$ whose values uniquely determine
those at all $r > r_b$. In particular they fix the bulk fields and
their derivatives at the other brane, and thereby dictate the
properties this brane must have to allow the geometry to be
properly completed.

Since our focus here is on formulation of the matching between the
bulk and the effective codimension-2 brane, we do not follow in
detail the properties of this second brane. We instead regard it
as being always adjusted as required if we desire to change the
properties of the $r=0$ brane in a particular way.

\section{Codimension-2 Actions and Matching}

In this section we use the $(d+1)$-dimensional codimension-1 brane
defined on the `cylinder', $\rho = \rho_b$, to define two kinds of
$d$-dimensional low-energy actions: the codimension-2 brane
action, $S_2$, appropriate to the description of the brane source
when the extra-dimensional bulk fields vary over scales much
larger than $\rho_b$; and the effective action, $S_{\rm eff}$,
describing brane physics at still-longer wavelengths, larger than
the size of the extra dimensions themselves.

Once these actions are defined, this section then recasts the
junction conditions to only refer to the codimension-2 quantities,
and to the properties of the bulk fields exterior to the brane.
This allows us to cast off the codimension-1 crutch by providing a
direct connection between the bulk configurations and the
properties of the effective codimension-2 objects which source
them.

\subsection{Low-energy interpretations for $t_{\mu\nu}$
and $t_{\theta\theta}$}
\label{effactionone}

We start by defining the regularized codimension-2 action, $S_2$,
and the very-low-energy action, $S_{\rm eff}$, and show how these
are well approximated by the dimensionally reduced stress
energies, $\ctwoTensn $ and $\ctwoPotl $, defined in
eqs.~\pref{T2defs} and \pref{U2defs}.

\subsubsection*{The codimension-2 brane action}

In the limit that a codimension-1 cylindrical brane has a very
small radius, it should admit an effective description as a
codimension-2 object. We define the action for this object by
dimensionally reducing the codimension-1 brane on its very small
circular direction.\footnote{We put aside for simplicity here a
more refined definition, based on multipole moments of the
microscopic codimension-1 brane, that can also allow the treatment
of sources that are not strictly axially symmetric.} One of the
results of this section is to show that the codimension-1 junction
conditions ensure that such a definition correctly reproduces the
proper scalar field properties near the brane.

In practice, for branes having small proper radius, this
dimensional reduction is well-approximated by a dimensional
truncation of the codimension-1 action's $\theta$ direction.
Writing $S_2 = \int \d^{d}x \; {\cal L}_2$ and $S_1 = \int
\d^{d+1}x \; {\cal L}_1$, we then find:
\be
    \L_2 = \int \d\theta \, \L_1
    = - \int \d\theta \sqrt{g_{\theta\theta}}
    \; \sqrt{-\cg} \left\{  \coneTensn (\phi) + \frac{n^2}{2}
    \, \coneKin (\phi) \; g^{\theta\theta} \right\} + \cdots\,,
\ee
where the ellipses denote corrections to the truncation
approximation. Using as before a trivial geometry for the interior
cap --- $e^{B(\rho_b)} = r_b$ and $W(\rho_b) = W_b$, where
$e^{-(d-1)W_b} = 1 + [(d-1)/4d] r_b^2 R$ --- then leads to
\be
    \L_2 = - \sqrt{-g} \; \ctwoTensn (\phi) \,,
\ee
with the codimension-2 tension, $\ctwoTensn $, as defined in
eq.~\pref{T2defs} and \pref{U2defs}. In terms of the useful
dimensionless quantities,
\be \label{UVDefs}
    \redoneTensn  = \kappa^2 r_b e^{dW_b} \, \coneTensn
    \quad \hbox{and} \quad
    \redoneKin  = \frac{\kappa^2 n^2 e^{dW_b}
    \coneKin }{2r_b} \,,
\ee
we have $\ctwoTensn = 2\pi ( \redoneTensn  + \redoneKin  )
/\kappa^2$.

\subsubsection*{Integrating out the bulk}

A second important low-energy quantity is the action, $S_{\rm
eff}$, relevant at energies below the KK scale, obtained by
completely integrating out all of the bulk degrees of freedom. It
is this action which is relevant to describing the physics seen by
brane-bound observers, including potential `low-energy'
applications to particle physics and cosmology. We here evaluate
this action at the classical level, where it is found by
eliminating the bulk fields from the microscopic action by
evaluating them at their classical solutions, regarded as
functions of the light fields, $\varphi^a$, that appear in the
low-energy theory: $\phi^{\rm cl} = \phi^{\rm cl}(\rho;\varphi)$.

When evaluated at the solution to Einstein's equations, the bulk
action becomes
\ba
    S_{EH}\left( \phi^{\rm cl}, g_{\ssM\ssN}^{\rm cl} \right)
    &=& - \frac{1}{2\kappa^2} \int \d^{D}x \, \sqrt{-g^{\rm cl}} \;
    \Bigl[ g_{\rm cl}^{\ssM\ssN} \Bigl( \cR^{\rm cl}_{\ssM\ssN}
    + \partial_\ssM \phi^{\rm cl}
    \, \partial_\ssN \phi^{\rm cl} \Bigr) + 2\kappa^2 V
    \Bigr] \nn\\
    &=& \frac{2}{d} \int \d^{D}x \, \sqrt{-g^{\rm cl}}
    \; V(\phi^{\rm cl}) \,,
\ea
and so vanishes completely for {\it any} solution if $V=0$.
Consequently, the total result for $S_{\rm eff}$ in the case of
vanishing\footnote{The corresponding argument for 6D chiral gauged
supergravity also gives a result completely localized at the brane
positions despite having a bulk potential, because $V$ also
cancels \cite{BdRHT}. The brane contribution in this case also
includes a contribution proportional to $\delta S_b/\delta \phi$.}
$V$ involves only fields evaluated at the brane positions:
\ba
    S_\eff(\varphi) &=& \left. S_{EH} + \sum_b (S_b + S_{GH})
    \right|_{\phi^{\rm cl}(\varphi)
    , g^{\rm cl}_{\ssM\ssN} (\varphi)} \nn\\
    &=& \frac{2}{d} \int \d^{D}x \, \sqrt{-g^{\rm cl}}
    \; V(\phi^{\rm cl}) +
    \sum_b \left\{ S_b - \frac{1}{\kappa^2}
    \int \d^{d+1} x \, \sqrt{-\hg} \; \Bigl[ K \Bigr]_b \right\}
    \nn\\
    &=& \frac{2}{d} \int \d^{D}x \, \sqrt{-g^{\rm cl}}
    \; V(\phi^{\rm cl}) + \sum_b \left\{ S_b - \frac{1}{d}
    \int \d^{d+1} x \, \sqrt{-\hg} \; \hg_{mn} t^{mn} \right\}
    \,,
\ea
where $S_b$ denotes the appropriate codimension-1 brane action,
and the sum is over all of the branes present in the geometry.
Here $S_{GH}$ denotes the standard Gibbons-Hawking action
\cite{GH} that is required for any codimension-1 brane that bounds
a bulk region, and the two terms in the `jump' form, $[K]_b$,
respectively arise from the bulk and the cap geometry interior to
each codimension-1 brane. The last equality follows from use of
the Israel junction conditions, eq.~\pref{IsraelJCs}.

Once this result is dimensionally reduced in the angular
directions, we obtain an effective lagrangian density, $\L_\eff$,
defined by $S_\eff \equiv \int \d^{d}x \, \L_\eff$, given as
\ba
    \L_\eff(\varphi) &=& \frac{2}{d} \int \d^{2}x \,
    \sqrt{-g^{\rm cl}}
    \; V(\phi^{\rm cl}) + \sum_b \int \d\theta \, \left\{ \L_{1b} -
    \frac{2}{d} \; \hg_{mn}
    \frac{\partial \L_{1b}}{\partial \hg_{mn}} \right\}  \,.
\ea
(The subscript `1' in this expression is meant to emphasize that
it is the codimension-1 brane action which is to be used.) In
particular, using
\be
    S_1 = - \int \d^{d+1}x \, \sqrt{-\hg} \,
    \left\{ \coneTensn  + \frac12 \, \coneKin
    \, \partial_m \sigma \,\partial^m \sigma
    \right\} \,,
\ee
for each brane, and specializing to $V = 0$ in the bulk, we find
$\L_\eff = - \sqrt{-g} \; U_\eff$, with
\be
    U_\eff = \frac{2\pi}{d} \sum_b
    \left\{ \frac{\redoneKin  - \redoneTensn  }{\kappa^2} \right\}_b
    = \sum_b \left( \frac{\ctwoPotl }{d} \right)_b \,.
\ee
Notice in particular that this last result shows that the
low-energy potential below the KK scale is governed by the
dimensionally reduced angular stress energy, $\ctwoPotl $, rather
than to the codimension-2 energy density, $\ctwoTensn $,
consistent with the known existence of bulk solutions sourced by
flat branes having nonzero tension.

When many branes are present, the low-energy action derived above
arises as a sum of functions of the dilaton, evaluated at the
position of a specific brane, $U_{{\rm eff}\,b} = U_{{\rm
eff}\,b}(\phi_{b}) = U_{{\rm eff}\,b}(\phi^{\rm cl}(\rho_{b}))$.
Consequently, each term in the sum has a different argument. These
all become related to one another through the bulk equations of
motion, however, and to understand the dynamics we are to express
each of these terms in terms of the light zero modes, $\varphi$,
which survive into the low-energy, $d$-dimensional, on-brane
theory (such as the constant mode of $\phi$, or the breathing mode
controlling the size of the extra dimensions). In principle,
because the symmetries that keep these modes light are broken by
the brane action, both can appear in $U_\eff$, and this provides
part of the dynamics which stabilizes their relative motion (or
allows them to run away from one another).

The interpretation of $U_2/d$ as a contribution to the on-brane
effective potential also provides useful information about the
relative sizes of the dimensionless quantities $r_b^2 R$,
$\kappa^2 \ctwoTensn$ and $\kappa^2 \ctwoPotl$, in the regime of
interest. It does so because the 4D Einstein equation ensures $R
\sim \kappa^2_d \ctwoPotl$, with the on-brane, $d$-dimensional
effective gravitational coupling given by $\kappa_d^2 \sim
\kappa^2/L^2$ where $L^2$ is a measure of the volume of the
geometry transverse to the branes. It follows from this that
\be
 r_b^2 R \sim \frac{r_b^2}{L^2} \, \kappa^2 U_\eff
 \ll \kappa^2 U_\eff \,,
\ee
and so generically our interest is for $r_b^2 R \ll \kappa^2
\ctwoPotl$, $\kappa^2 \ctwoTensn$. Furthermore, validity of the
semiclassical techniques we use also requires both $\kappa^2
\ctwoPotl$ and $\kappa^2 \ctwoTensn$ be small compared to unity.

\subsection{Matching and the codimension-2 action}
\label{effactiontwo}

Recall that our goal is to relate the integration constants that
appear in the bulk classical solutions directly to the properties
of the codimension-2 brane. Junction conditions like
eqs.~\pref{dilatonJC} through \pref{IsraelJCthetatheta} are
unsatisfying in this regard, since they instead relate the bulk to
the properties of the codimension-1 brane action and to the
geometry of the capped interior. We extend these jump conditions
to directly involve codimension-2 quantities in the present
section.

\subsubsection*{Codimension-2 Action and Bulk Derivatives}

The first step is accomplished by multiplying the jump conditions
through by $e^{B+d W}$. For the dilaton condition, using $\left[
e^{B+dW} \partial_\rho \right]_b = \left[ \xi (r-l) \partial_r
\right]_b = \left[ r \partial_r \right]_b$ --- where the last
equality is only true at $r=r_b$ --- gives
\be \label{dilatonJC2}
    \Bigl[ r\, \partial_r \phi \Bigr]_b =
    \Bigl[ e^{B+d W} \, \partial_\rho \phi \Bigr]_b
    = \frac{\kappa^2 \ctwoTensn '}{2\pi}  \,,
\ee
where $\ctwoTensn ' = \partial \ctwoTensn /\partial\phi_b$. The
$(\mu\nu)$ and $(\theta\theta)$ Israel junction conditions
similarly become
\ba \label{IsraelJCmunu2}
    \Bigl[ (d-1) r \, \partial_r W + r \, \partial_r B
    \Bigr]_b &=& \Bigl[ e^{B+d W} \, \Bigl(
    (d-1)\partial_\rho W + \partial_\rho B \Bigr) \Bigr]_b
    = - \frac{\kappa^2 \ctwoTensn }{2\pi} \\
\label{IsraelJCthetatheta2}
    \hbox{and} \qquad
    \Bigl[ d\, r \, \partial_r W \Bigr]_b
    &=& \Bigl[ d e^{B+dW} \,\partial_\rho W \Bigr]_b
    = \frac{\kappa^2 \ctwoPotl }{2\pi}   \,.
\ea

Next we remove all reference to the interior geometry by using its
explicit properties, and so it is at this point that we assume
that $V$ may be neglected inside the cap (and so, by continuity,
also for the external solution nearby the brane). We then find:
$r\partial_r \phi_i = 0$, $(d-1)r \partial_r W_i = -2 \, r^2/(r^2
+ l_{\ssW i}^2)$ and $r \partial_r B_i = 1$, with $(d-1)l_{\ssW
i}^2 = 4d/R$. This leads to the results
\ba \label{BulkFieldAsymptotics}
    \Bigl( e^{B_e+dW_e} \partial_\rho\phi_e \Bigr)_{\rho\to \rho_b}
    &=& \frac{\kappa^2
    \ctwoTensn ' (\phi_b)}{2\pi} \\
    \Bigl( e^{B_e + dW_e} \partial_\rho W_e
    \Bigr)_{\rho\to\rho_b}
    &=& - \frac{2 r_b^2}{(d-1)(r_b^2 + l_{\ssW i}^2)} +
    \frac{\kappa^2 \ctwoPotl (\phi_b)}{2\pi d} \nn\\
    &=& - \frac{2 \,r_b^2 R}{(d-1) \,r_b^2 R + 4d} +
    \frac{\kappa^2 \ctwoPotl (\phi_b)}{2\pi d} \\
    &\simeq&
    \frac{\kappa^2 \ctwoPotl (\phi_b)}{2\pi d} \nn\\
    \label{BulkFieldAsymptoticsa}
  \Bigl( e^{B_e + dW_e} \partial_\rho B_e \Bigr)_{\rho
    \to \rho_b}
    &=& 1 - \frac{\kappa^2}{2\pi} \left[ \ctwoTensn (\phi_b)
    + \left( \frac{d-1}{d} \right) \ctwoPotl (\phi_b) \right]\,,
    \label{BulkFieldAsymptoticsb}
\ea
which provides the desired relation between the codimension-2
brane action and the radial near-brane derivatives of bulk fields
in the exterior geometry. Notice in particular that the first of
these equations shows how it is the derivative of $\ctwoTensn $
that governs the radial gradient of the dilaton, in precisely the
way one would naively expect for a $\delta$-function localized
codimension-2 source.

\subsubsection*{Matching of Bulk Integration Constants}

In principle, these last equations allow the determination of some
of the bulk integration constants in terms of source brane
properties. For instance, if the exterior bulk geometry near a
specific brane has the form of the exact solutions given in
eqs.~\pref{generalsolnshft}, then the left hand sides may be
explicitly evaluated using eqs.~\pref{genderivshft} and
\pref{genderivshftW},
\be \label{BulkFieldAsymptotics2}
    \xi p_\phi = \frac{\kappa^2 \ctwoTensn ' }{2\pi} \,, \qquad
    \xi p_\ssB = 1 - \frac{\kappa^2}{2\pi}
    \left[ \ctwoTensn  + \left( \frac{d-1}{d} \right) \ctwoPotl
    \right]\,, \nn
\ee
and
\be \label{BulkFieldAsym3}
    - \frac{\xi}{d-1} \left\{ p_\ssB + \Omega \, \left[
    \frac{(r_b -l)^{2\Omega}
    - l_\ssW^{2\Omega}}{(r_b-l)^{2\Omega}
    + l_\ssW^{2\Omega}} \right] \right\}
    = - \frac{2\, r_b^2 R}{(d-1)\,r_b^2 R + 4d} +
    \frac{\kappa^2 \ctwoPotl }{2\pi d} \,.
\ee
Neglecting, to first approximation, $r_b^2 R$ relative to
$\kappa^2 \ctwoTensn$ and $\kappa^2 \ctwoPotl$, this last
condition simplifies to
\be \label{BulkFieldAsym3app}
    \xi (\Omega -  p_\ssB)
    \simeq
    \frac{\kappa^2}{2\pi} \left( \frac{d-1}{d} \right) \ctwoPotl \,.
\ee

We imagine solving these constraints for three of the as-yet
unchosen parameters $p_\phi$, $p_\ssB$, $r_b$ and $\phi_b$, given
assumptions for the underlying brane coupling functions
$\ctwoTensn (\phi_b,r_b)$ and $\ctwoPotl (\phi_b,r_b)$. For
instance, eqs.~\pref{BulkFieldAsymptotics2} directly give $p_\phi$
and $p_\ssB$ as functions of $\phi_b$ and $r_b$. Using these in
eq.~\pref{BulkFieldAsym3} or \pref{BulkFieldAsym3app} then gives a
condition relating $\phi_b$ to $r_b$.

This last condition is conceptually important, because it allows
the variable $r_b$ to be eliminated from the codimension-2 tension
and potential, thereby allowing these to be expressed purely in
terms of $\phi_b$ (and, possibly, geometric quantities like $R$
that characterize the bulk). That is, it allows us to trade the
functions of two variables, $\ctwoTensn(\phi_b,r_b)$ and
$\ctwoPotl(\phi_b,r_b)$, given by eqs.~\pref{T2defs} and
\pref{U2defs}, with
\be
 \ctwoTensn(\phi_b) := \ctwoTensn(\phi_b, r_b(\phi_b))
 \qquad \hbox{and} \qquad
 \ctwoPotl(\phi_b) := \ctwoPotl(\phi_b, r_b(\phi_b)) \,.
\ee

The explicit calculation of $r_b(\phi_b)$ using
eq.~\pref{BulkFieldAsym3app} simplifies considerably once we use
the weak-gravity limits $\kappa^2 \ctwoTensn /2\pi \ll 1$ and
$\kappa^2 \ctwoPotl /2\pi \ll 1$, that underlie our entire
semiclassical analysis. The simplification comes because these
imply $p_\phi$ and $\delta p_\ssB = p_\ssB - 1$ are both small, in
which case $\Omega \simeq p_\ssB + \frac{1}{2d} (d-1)
(p_\phi^2/p_\ssB) + {\cal O}(p_\phi^4)$.

Because of its conceptual importance, rather than directly
exploring its solution immediately, we first pause to re-derive
eq.~\pref{BulkFieldAsym3app} in a way which does not rely on the
explicit form of specific solutions to the bulk field equations,
and so which also includes the situation where the bulk potential,
$V$, does not vanish. Once re-derived in this way we explore its
consequences for in an explicit example.

\subsubsection*{Curvature Constraint}

The relation we seek can be identified very robustly because it
expresses the `Hamiltonian' constraint for integrating the field
equations in the $\rho$ direction. As such it can be regarded as a
restriction on the form of the brane action that must be satisfied
in order for there to be maximally symmetric and axially symmetric
solutions having a given brane curvature, $R$.\footnote{This
constraint was derived in ref.~\cite{SN}, but interpreted somewhat
differently.}

To derive this constraint we first eliminate the second
derivatives, $W''$ and $B''$, from the bulk Einstein equations by
taking the combination [$d(\mu\nu) - (\rho\rho) +
(\theta\theta)$], and then use $\xi (r - l) \partial_r = e^{B+dW}
\partial_\rho$, with the result
\be \label{bulkconstraint}
    d [(r-l) \partial_r W] \Bigl\{ (d-1) [(r-l) \partial_r W] + 2
    [(r-l)\partial_r B] \Bigr\}
    - [(r-l) \partial_r \phi]^2 +
    \frac{1}{\xi^2} e^{2[B+dW]} \Bigl( \cR + 2\kappa^2 V \Bigr) = 0 \,.
\ee
Next, take the limit of this equation as $r \to r_b$, approaching
from the exterior side, and use $\xi (r-l) \partial_r \to r_b
\partial_r$ as well as eqs.~\pref{BulkFieldAsymptotics} to evaluate the
derivatives of $W_e$, $B_e$ and $\phi_e$ in this limit. Finally,
using $e^{B_b} = r_b$ and $\cR = R e^{-2W}$, a bit of algebra
gives the following brane constraint
\ba \label{braneconstraint}
    &&2d \, \psi_b \, \redtwoTensn  +
    \redtwoPotl  \left\{ 2 - 2\,\redtwoTensn  - \left( \frac{d-1}{d}
    \right) \redtwoPotl  \right\} - \left( \redtwoTensn ' \right)^2 \nn\\
    && \qquad\qquad\qquad
    + d\psi_b\Bigl[ (d-1) \psi_b -2  \Bigr]
    + r_b^2 e^{2dW_b}\Bigl( R e^{-2W_b} + 2\kappa^2 V_b
    \Bigr) = 0 \,,
\ea
where $V_b = V(\phi_b) = V(\phi(\rho_b))$, while $\redtwoTensn  :=
\kappa^2 \ctwoTensn /2\pi = \redoneTensn  + \redoneKin $ and
$\redtwoPotl := \kappa^2 \ctwoPotl /2\pi = \redoneKin  -
\redoneTensn $ are convenient dimensionless measures of the
codimension-2 brane actions. The quantity $e^{W_b} = e^{W_e(r_b)}$
is given, as above, by
\be
    e^{W_b} = \left( \frac{l_{\ssW i}^2}{r_b^2 + l_{\ssW i}^2}
    \right)^{1/(d-1)} = \left( 1 + \frac{d-1}{4d} \,r_b^2 R
    \right)^{-1/(d-1)} \,,
\ee
while $\psi_b$ is defined as the combination
\be
  \psi_b = \frac{2\, r_b^2 R}{(d-1) \,r_b^2 R + 4d}
  = \frac{1}{2d} \, e^{(d-1)W_b} \, r_b^2 R \,.
\ee

Equation \pref{braneconstraint} directly relates the brane
curvature to the amount of matter on the brane, and when written
in terms of the Hubble scale, $R \propto H^2$, for an FRW
foliation of a de Sitter or anti-de Sitter geometry, it provides a
generalization to codimension-2 branes of the much-studied
codimension-1 brane-world modification to Hubble's law.

However, in the small-brane regime we may neglect the small
quantities $r_b^2 R$ and $r_b^2 \kappa^2 V_b$ in
eq.~\pref{braneconstraint}, leading to the following expression:
\be \label{braneconstraintexp}
    \redtwoPotl  \left\{ 2 - 2\,\redtwoTensn  - \left( \frac{d-1}{d}
    \right) \redtwoPotl  \right\} - \left( {\redtwoTensn }' \right)^2
    \simeq 0 \,,
\ee
which clearly can be used to learn $\ctwoPotl$ from $\ctwoTensn$
or vice versa. Eq.~\pref{braneconstraintexp} provides the desired
generalization of eq.~\pref{BulkFieldAsym3app} to the case where
$V \ne 0$, and so where the explicit form of the bulk solutions is
not known.

The solutions for $r_b(\phi_b)$ obtained by solving
eq.~\pref{braneconstraintexp} can be found explicitly by expanding
in powers of the small quantities $\redtwoPotl = \kappa^2
\ctwoPotl/2\pi$ and $\redtwoTensn = \kappa^2 \ctwoTensn/2\pi$.
Writing $r_b = r_{b0} + \delta r$, we see that the leading
contribution satisfies ${\ctwoPotl}_0(\phi) := \ctwoPotl
(\phi,r_{b0}) \simeq 0$, and so gives $r_{b0}(\phi_b)$ as
\be
  r_{b0}(\phi) = |n| \sqrt{ \frac{\coneKin(\phi)}{2
  \, \coneTensn(\phi)}} \,.
\ee
This makes the leading form for the tension become
\be
 {\ctwoTensn}_0(\phi) \simeq \ctwoTensn(\phi, r_{b0}(\phi))
 = 2\pi |n|\sqrt{2 \, \coneTensn \coneKin}  \,.
\ee

Working to next order gives the following, leading condition for
$\delta r$:
\be
 2 \left( \frac{\partial\redtwoPotl }{\partial r_b} \right)_0
 \delta r - \left( {{\redtwoTensn}_0} ' \right)^2
 \simeq 0 \,,
\ee
where $({\partial \redtwoPotl/\partial r_b})_0 = -2\kappa^2 T_1$
and ${{\redtwoTensn}_0}' = |n| \kappa^2 (Z_1 T_1)' /\sqrt{2 \, T_1
Z_1} = \kappa^2[ r_{b0} T_1' + (n^2/2r_{b0}) Z_1']$. Consequently
\be
 \delta r \simeq - \frac{n^2 \kappa^2
 [(T_1 Z_1)']^2}{8 T_1^2 Z_1} = - \frac{r_{b0}^2 \kappa^2
 [(T_1 Z_1)']^2}{4 T_1 Z_1^2}
 \,,
\ee
and so the leading contribution to the on-brane potential becomes
\be
 \ctwoPotl(\phi) \simeq \left( \frac{\partial \ctwoPotl}{
 \partial r_b} \right)_0 \delta r
 = \frac{\kappa^2}{4\pi} \left( {{\ctwoTensn}_0}' \right)^2
 = \left( \frac{\pi \kappa^2 n^2}{2} \right)
 \frac{[(T_1 Z_1)']^2}{T_1 Z_1} \,.
\ee

\subsection{An Example}

To make all this perfectly concrete consider a brane for which
\be
 \coneTensn (\phi_b) = A_\ssT e^{- t \phi_b}
 \qquad \hbox{and} \qquad
 \coneKin(\phi_b) = A_\ssZ e^{-z \phi_b} \,,
\ee
and so
\be
 \ctwoTensn(\phi_b,r_b) \simeq 2\pi \left[ r_b A_\ssT e^{-t \phi_b}
 + \left( \frac{n^2 A_\ssZ}{2r_b} \right)
 e^{-z \phi_b} \right]\,,
\ee
and
\be
 \ctwoPotl(\phi_b,r_b) \simeq -2\pi \left[ r_b A_\ssT e^{-t \phi_b}
 - \left( \frac{n^2 A_\ssZ}{2r_b} \right) e^{-z \phi_b}
 \right] \,.
\ee

In this case the zeroth-order brane size is
\be
 r_{b0} = |n| \sqrt{\frac{A_\ssZ}{2 A_\ssT}} \,
 e^{-(z-t)\phi_b/2} \,,
\ee
with ${\cal O}(\kappa^2)$ correction
\be
 \delta r \simeq - \frac{n^2 \kappa^2
 [(T_1 Z_1)']^2}{8 T_1^2 Z_1} = - \frac18 \, n^2 (t+z)^2
 \kappa^2 A_\ssZ e^{-z \phi_b} \,.
\ee
Using these the leading contribution to the codimension-2 brane
tension and on-brane potential then become
\be
 \ctwoTensn(\phi_b) \simeq {\ctwoTensn}_0(\phi_b)
 = 2 \pi |n| \sqrt{2A_\ssT A_\ssZ}
 \, e^{-(t+z) \phi_b/2} \,,
\ee
and
\be
 \ctwoPotl(\phi_b) \simeq
 \frac{\kappa^2}{4\pi} \left({{\ctwoTensn}_0}' \right)^2
 = \frac{\pi}{2} \, n^2 (t+z)^2 \kappa^2
 A_\ssT A_\ssZ e^{-(t+z) \phi_b}
 \,.
\ee
The powers $p_\phi$ and $p_\ssB$ then are
\ba
    \xi p_\phi &=& - \frac{|n|}{2} \, (t+z) \kappa^2
    \sqrt{ 2 A_\ssT A_\ssZ} \;
    e^{-(t+z) \phi_b/2} \,, \nn\\
    \xi p_\ssB &=& 1 - |n| \kappa^2 \sqrt{2 A_\ssT A_\ssZ}
    \; e^{-(t+z)\phi_b/2} + {\cal O}(\kappa^4 )
    \,.
\ea

Clearly these expressions show that special things happen when
$t+z = 0$, as should be expected given that this is the choice
that preserves one combination of the symmetries ---
eqs.~\pref{axionsymmetry} and \pref{scalesymmetry} --- that the
bulk equations enjoy when $V=0$.

\section{Renormalized Brane Actions}

In many ways the previous section solves the problem of relating
bulk properties to those of the codimension-2 branes that source
them, by giving an explicit connection between asymptotic
near-brane derivatives of bulk fields and the codimension-2 brane
action, $\ctwoTensn$, and on-brane potential, $\ctwoPotl$. An
important drawback is its explicit dependence on fields (like
$\phi_b$) evaluated at the microscopic scale, $r_b$, which
characterizes the size of the codimension-1 crutch. This is a
drawback inasmuch as one would like to take microscopic quantities
like $r_b$ and $l$ to zero when describing macroscopic physics on
much larger scales, and the bulk fields generically diverge in
this limit. For instance, relative to $\phi_0 = \phi(r_0)$
evaluated in the bulk, we have $\phi_b = \phi_0 + p_\phi
\ln\left[(r_b - l)/(r_0 - l) \right]$, which diverges
logarithmically (when $p_\phi \ne 0$) as $r_b, l \to 0$. This
makes the limit of a microscopic codimension-2 brane slightly more
subtle than is generally encountered in codimension-1
applications.

This section shows how to address this limit, and the idea is
simple: we express the matching conditions in terms of a
`renormalized' codimension-2 brane action whose brane couplings
are independent of the value of `regularization' scale, $r_b$,
ensuring that the limit $r_b\to 0$ does not introduce divergences.
Such a {\it classical} renormalization of effective codimension-2
brane couplings has already been applied elsewhere \cite{GW,CdR},
although earlier authors typically rely on graphical methods near
flat space. Our aim here is to show that these results for the
classical renormalizations can be extended to include nontrivial
bulk fields by a very simple modification of the junction
conditions discussed above, together with simple geometrical
considerations. Our formalism reproduces in appropriate limits
earlier calculations of the running of classically renormalized
couplings, without the need for graphical calculations.

The idea is to define a `renormalized' codimension-2 brane action,
$\overline S_2$, in a way that is formally very similar to the
`regularized' action, $S_2$, used heretofore. However, rather than
defining this action in terms of a regularizing codimension-1
brane at $r = r_b$ as in previous sections, we instead similarly
define $\overline S_2$ at a much larger, floating, radius $r =
\bar r$, at which a fictitious codimension-1 brane is imagined to
be located. We {\it define} the action of this brane to be
whatever is required to source precisely the same bulk fields as
are produced by the much smaller regularized brane, described by
$S_2$. We shall find that $\overline S_2$ defined in this way
makes no reference to the microscopic scale, $r_b$, and so remains
well-defined if $r_b$ is taken to zero. Furthermore, since the
scale, $\bar r$, at which the renormalized action is defined is
completely arbitrary, nothing physical can depend on it. This
condition allows the derivation of renormalization-group (RG)
conditions for the action $\overline S_2$, that we show reduce to
those derived by earlier workers in the appropriate limits.

\EPSFIGURE[t]{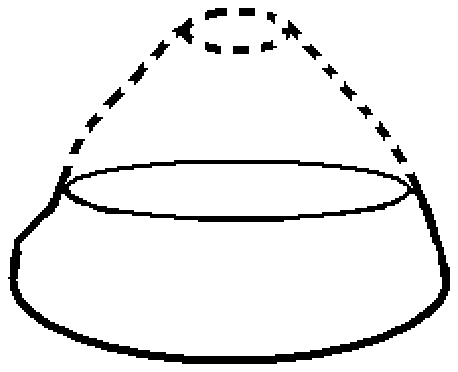, width = 0.4\textwidth,
height=5cm}{\sl A cartoon of the exterior geometry cut off by a
larger `floating' brane.\label{fig:rencap}}

\subsection{Floating Branes}

To this end, consider the bulk fields sourced by a codimension-2
brane, which we imagine is regularized by a codimension-1 brane
situated at $r = r_b$, as before. Now, imagine drawing a large,
fictitious circle at a much larger radius $\bar r \gg r_b$, but
which is nevertheless much smaller than the typical scale (such as
$l_\ssW$) defined by the bulk geometry. We place a fictitious
codimension-1 `floating' brane (and, by dimensional reduction, an
implicit effective codimension-2 brane) at $\bar r$, and replace
the full geometry for $r < \bar r$ by a nonsingular cap geometry.
As before, we ask this interior geometry to match continuously to
the exterior solution at $r = \bar r$, but with the important
difference that this time we use these conditions to fix
integration constants in the interior solution, with the exterior
geometry regarded as given (rather than the other way around, as
before).

When $V$ is negligible near the brane we use precisely the same
exterior solution as before, eqs.~\pref{generalsolnshft} and
\pref{generalsolnWshft}, and so find the following values at $r =
\bar r$:
\be \label{extrensoln1}
 e^{\bar\phi} = e^{\phi_b} \left( \frac{\bar r - l}{r_b - l}
 \right)^{p_\phi} , \qquad
 e^{\bar B} = r_b \left( \frac{\bar r - l}{r_b
 - l} \right)^{p_\ssB}
 \,,
\ee
and
\be \label{extrensoln2}
 e^{(d-1) \overline W} = \frac{ \left[ (r_b - l)/l_\ssW
 \right]^\Omega + \left[ l_\ssW/(r_b - l) \right]^\Omega}{
 \left[ (\bar r-l)/ l_\ssW \right]^\Omega + \left[ l_\ssW /
 (\bar r-l) \right]^\Omega} \left( \frac{r_b - l}{\bar r - l}
 \right)^{p_\ssB} e^{(d-1) W_b} \,.
\ee
The goal is to repeat the arguments of the previous sections to
express the near-brane derivatives in terms of an action defined
at $\bar r$ rather than $r_b$. This is useful because in any limit
where $r_b$ and $l$ are taken to zero, the quantities $\bar \phi$,
$\bar B$ and $\overline W$ will be held constant.

Continuity and regularity at the potential singularity at $r=0$
require the interior `floating' solution (for negligible $V$) to
become
\be \label{interiorbar}
 \phi_f = \bar\phi \,, \quad
 e^{B_f} = \left( \frac{r}{\bar r} \right)^p
 e^{\bar B}
 \quad \hbox{and} \quad
 e^{(d-1)W_f} = \left[ \frac{\bar r^{2p} + l_{\ssW f}^{2p}}{
 r^{2p} + l_{\ssW f}^{2p}} \right] \,
 e^{(d-1) \overline W} \,,
\ee
where $p$ is chosen to ensure the geometry has no conical defects.
To determine what this requires we write the proper distance
within the cap (see appendix \ref{App:SolvingFieldEq}) as
\be
 \exd \rho = e^{B_f + d W_f} \, \exd \ln r
 = e^{\bar B + d \overline W} \, \left( \frac{r}{\bar r} \right)^p
  \left[ \frac{\bar r^{2p} + l_{\ssW f}^{2p}}{
 r^{2p} + l_{\ssW f}^{2p}} \right]^{d/(d-1)}
 \frac{\exd r}{r}\,,
\ee
where $\xi_f = 1$ is chosen to ensure continuity of $\exd
\rho/\exd r$ at $r = \bar r$. In terms of $\rho$ we have $e^{B_f}
= \alpha_f \, \rho + {\cal O}(\rho^3)$, with
\be
 \alpha_f = p \, e^{- d \overline W} \,
  \left[1 + \left( \frac{\bar r}{ l_{\ssW f}}
  \right)^{2p} \right]^{-d/(d-1)}  \,,
\ee
and so to avoid a conical singularity we choose
\be \label{psoln}
 p = e^{d\overline W} \left[ 1 + \left( \frac{\bar r^2}{
 l_{\ssW f}} \right)^{2p} \right]^{d/(d-1)} > 0 \,.
\ee
Notice that, unlike for the regularized brane, $W_f$ and
$e^{B_f}$ need not vanish at the same place. As before, the
constant $l_{\ssW f}$ is set by continuity of the on-brane
curvature, with
\be \label{Rfsoln}
 R = \frac{4d\, p^2}{(d-1) \bar r^2} \left( \frac{\bar r}{
 l_{\ssW f}} \right)^{2p} \left[ 1
 + \left( \frac{\bar r}{l_{\ssW
 f}} \right)^{2p} \right]^{-2} e^{-2(d-1) \overline W}
 = \frac{4d \, p^{2/d}}{(d-1) \bar r^2} \left( \frac{\bar r}{
 l_{\ssW f}} \right)^{2p} \,.
\ee

Turning to the jump conditions across $r = \bar r$, we come to the
main point: we {\it define} the brane action at $\bar r$ by the
condition that it produce the required discontinuity in the bulk
field derivatives. That is,\footnote{In a spirit similar to
ref.~\cite{RSRG}.} we now regard eqs.~\pref{BulkFieldAsymptotics}
-- \pref{BulkFieldAsymptoticsb} (reproduced again here)
\ba \label{dilatonJCbar}
    \frac{\kappa^2 {\renctwoTensn} ' (\overline \phi)}{2\pi} &=&
   \left( e^{ \overline B_e + d \overline W_e}
 \partial_{ \rho} \phi_e \right)_{\rho = \overline \rho} \\
\label{JCdilatonbar}
    \frac{\kappa^2 \renctwoTensn (\overline \phi)}{2\pi} &\simeq&
    1 - \left( e^{ \overline B_e+d \overline W_e} \,
   \Bigl[ (d-1)  \partial_{ \rho} W_e + \partial_{ \rho} B_e
   \Bigr] \right)_{\rho = \overline \rho} \\
\label{IsraelJCthetathetabar}
    \frac{\kappa^2 \renctwoPotl (\overline \phi)}{2\pi}  &\simeq&
   \left( d\,  e^{\overline B_e + d\overline W_e}
    \partial_{ \rho} W_e \right)_{\rho = \overline \rho}
\ea
as being solved for the effective actions, $\renctwoTensn $ and
$\renctwoPotl $, given the known external bulk profiles sourced by
the underlying regularized brane defined at $r = r_b$ (together
with the singularity-free internal profiles they match across to
at $r = \overline r$). The approximate equalities in these
equations indicate the neglect of $\bar r^2 R \propto (\bar r /
l_{\ssW f})^{2p}$, as was also done in earlier sections when
neglecting $r_b^2 R$ in eqs.~\pref{BulkFieldAsymptotics} to
\pref{BulkFieldAsymptoticsb}.

One might worry that the three conditions,
eqs.~\pref{dilatonJCbar} through \pref{IsraelJCthetathetabar},
might overdetermine the two functions $\renctwoTensn$ and
$\renctwoPotl$, however this does not happen ultimately because
these equations are related to one another by the bulk field
equations and Bianchi identities. In fact, since any solution is
required to satisfy the curvature constraint --- {\it c.f.}
eq.~\pref{braneconstraintexp},
\be \label{braneconstraintren}
    \overline \redtwoPotl  \left\{ 2 - 2\, \overline
    \redtwoTensn  - \left( \frac{d-1}{d}\right)
    \overline \redtwoPotl  \right\} - \left( {\overline
    \redtwoTensn} ' \right)^2
    \simeq 0 \,,
\ee
this provides the most efficient means for finding $\renctwoPotl$
given $\renctwoTensn$, and vice versa, where $ \overline
\redtwoTensn = \kappa^2 \renctwoTensn / 2\pi$ and $ \overline
\redtwoPotl = \kappa^2 \renctwoPotl /2\pi$.

\subsubsection*{Renormalized actions and near-brane asymptotics}

Once the renormalized action is constructed in this way, it can be
related to the integration constants of the bulk solutions. For
instance, if we assume solutions are given by
eqs.~\pref{extrensoln1} and \pref{extrensoln2}, then the constants
$\xi$, $p_\phi$ and $p_\ssB$ are directly related to the
renormalized action by
\be \label{RGconst}
    \xi p_\phi = \frac{\kappa^2 \renctwoTensn '}{2\pi}
    \quad \hbox{and} \quad
    \xi p_\ssB = 1 - \frac{\kappa^2}{2\pi} \left[
    \renctwoTensn + \left( \frac{d-1}{d} \right) \renctwoPotl
    \right] \,.
\ee
The main difference between these and earlier formulae comes from
the observation that their right-hand sides remain finite as $r_b,
l \to 0$ with $\bar r$ and $\xi = [1-(l/r_b)]^{-1}$ fixed. We may
accordingly use in them $l_\ssW \gg \bar r \gg |l|$, while we
earlier had $l_\ssW \gg r_b \simeq |l|$.

\subsection{Codimension-2 RG Flow}

Rather than directly solving the above equations, it is often
simpler instead to obtain the renormalized action by solving an
appropriate renormalization group (RG) equation. In this section
we derive such an equation for the floating brane action, using
the bulk field equations and brane junction conditions. We then
examine in detail their form for a special case already studied in
the literature \cite{GW} using perturbative methods, reproducing
the previous results and extending them taking into account the
coupling with gravity.

\subsubsection*{Derivation of the RG equations}

Since the position of the floating brane, $\overline r$, is
completely arbitrary, physical quantities do not depend on it.
This is true in particular for the bulk field profiles themselves,
since the floating brane tension, $\renctwoTensn$, and on-brane
potential, $\renctwoPotl$, are defined to vary with $\bar r$ in
precisely the way required to leave bulk field profiles unchanged.
This observation provides an alternative way to derive these
renormalized quantities: by setting up and solving the
differential conditions that express the independence of the bulk
fields to changes in $\bar r$. The resulting equations are RG
equations inasmuch as they express the independence of quantities
under changes to $\bar r$, in much the same way as more
traditional RG equations express the independence of physical
quantities to the arbitrary renormalization point, $\mu$.

Since the relationship between the brane action and the bulk
fields is dictated by the field equations themselves, we derive
the floating equations by using the junction conditions after
directly applying the differential operator
\be
    {\cal D} = e^{\overline B + d \overline W}
    \frac{\partial}{\partial \overline \rho}
    \,,
\ee
to the renormalized actions, where it is understood that all the
integration constants in the {\em external} bulk fields are held
fixed when doing so. This means that ${\cal D}$ agrees with
$e^{B_e + d W_e} \partial_\rho = \xi (r - l) \partial_r$ when
applied to external bulk fields, with $r$ then taken to $\bar r$.
The same need not be true for the interior solutions, since
--- {\it c.f.} eqs.~\pref{interiorbar} --- these have integration
constants that depend explicitly on $\bar r$ (and so on $\bar
\rho$). To derive the RG equation we therefore apply ${\cal D}$ to
the junction conditions, eqs.~\pref{dilatonJC2} through
\pref{IsraelJCthetatheta2}, and simplify the result using the bulk
field equations.

For example, applying ${\cal D}$ to eq.~\pref{dilatonJC2} gives
\ba \label{RGU2'a}
   {\cal D} \; \frac{\kappa^2 {\renctwoTensn}'}{2\pi}
   &=& \left( e^{B_e+dW_e} \partial_{\rho} \Bigl[ e^{B_e+dW_e}
    \partial_{\rho} \phi_e
    \Bigr] \right)_{\rho = \overline \rho}
    - \left( e^{\bar B+d \overline W} \partial_{\bar \rho}
    \Bigl[ e^{B_f+dW_f} \partial_{\rho} \phi_f
    \Bigr]_{\rho = \overline \rho}  \right) \nn\\
    &=& \Bigl[ e^{B+dW} \partial_{\rho} \Bigl( e^{B+dW}
    \partial_{\rho} \phi \Bigr) \Bigr]_b
    - \left( e^{\bar B + d \overline W} \partial_{\bar \rho}
    \alpha^a \right) \left[ \frac{\partial}{\partial \alpha^a}
    \left( e^{B_f + d W_f} \partial_\rho \phi_f
    \right) \right]_{\rho = \bar \rho} \nn\\
    &=& - \left( e^{B_f + d W_f} \partial_{\bar \rho}
    \alpha^a \right) \left[ \frac{\partial}{\partial \alpha^a}
    \left( e^{B_f + d W_f} \partial_\rho \phi_f
    \right) \right]_{\rho = \bar \rho} \,,
\ea
where, as before, $\Bigl[X \Bigr]_b$ denotes the jump of the
quantity $X$ across $\rho = \bar \rho$, and the $\alpha^a$
collectively denote the integration constants of the interior
solution. The last equality then uses the dilaton field equation,
eq.~\pref{bulkrhoeqns}, to write the discontinuity as $\Bigl[
e^{2(B+dW)}\kappa^2 V' \Bigr]_b$, which vanishes because of the
continuity of $\phi$ and $V$ across the brane. When $V$ is
negligible in the near-brane limit, the right-hand-side of the
last equality in eq.~\pref{RGU2'a} can be evaluated explicitly
using the known cap solutions, giving
\be \label{RGU2'}
   {\cal D} \; \frac{\kappa^2 {\renctwoTensn}'}{2\pi}
   = 0 \,,
\ee
because $\partial_\rho \phi_f = 0$.

Similarly, applying ${\cal D}$ to eq.~\pref{IsraelJCthetatheta2}
and using the $(\mu\nu)$ Einstein equation of
eq.~\pref{bulkrhoeqns} gives
\ba \label{RGT2}
   {\cal D} \; \frac{\kappa^2 {\renctwoPotl}}{2\pi}
   &=& \Bigl[ e^{B+dW} \partial_{\rho} \Bigl( e^{B+dW}
    \partial_{\rho} W \Bigr) \Bigr]_b
    - \left( e^{\bar B + d \overline W} \partial_{\bar \rho}
    \alpha^a \right) \left[ \frac{\partial}{\partial\alpha^a}
    \left( e^{B_f + d W_f} \partial_\rho W_f
    \right) \right]_{\rho = \bar \rho} \nn\\
    &=& - \left( e^{\bar B + d \overline W} \partial_{\bar \rho}
    \alpha^a \right) \left[ \frac{\partial}{\partial\alpha^a}
    \left( e^{B_f + d W_f} \partial_\rho W_f
    \right) \right]_{\rho = \bar \rho}  \,,
\ea
which uses the continuity of $e^{2(B+dW)}[ \cR + 2\kappa^2 V]$
across $r = \bar r$. Finally, applying ${\cal D}$ to
eq.~\pref{IsraelJCmunu2} and using the $(\mu\nu)$ and
$(\theta\theta)$ equations of \pref{bulkrhoeqns} implies
\be \label{RGU2}
   {\cal D} \; \frac{\kappa^2 {\renctwoTensn}}{2\pi}
   = \left( e^{\bar B + d \overline W} \partial_{\bar \rho}
   \alpha^a \right) \left[ \frac{\partial}{\partial \alpha^a}
    \left( e^{B_f+dW_f} \partial_{\rho}
    \Bigl[ (d-1) W_f + B_f \Bigr]
    \right) \right]_{\rho = \overline \rho}
    \,,
\ee
which uses continuity of $e^{2 (B + d W_e)} \left[ \left(
{(d-1)}/{d} \right) \cR + 2\kappa^2 V \right]$.

When $V$ is negligible near the brane (and so also inside the cap)
we can evaluate the relevant derivatives explicitly, using $e^{B_f
+ dW_f} \partial_\rho X_f = r \partial_r X_f$ with
\be
  r  \partial_r B_f = p
 \quad\hbox{and} \quad
 r  \partial_r W_f = - \frac{2p}{d-1}
 \left( \frac{r^{2p}}{r^{2p} + l_{\ssW f}^{2p}}
 \right)\,,
\ee
and so
\be
 \left( e^{\bar B + d \overline W} \partial_{\bar \rho}
   \alpha^a \right) \left[ \frac{\partial}{\partial \alpha^a}
    \left( e^{B_f+dW_f} \partial_\rho B_f \Bigr]
    \right) \right]_{\rho = \overline \rho}
 = \bar r \partial_{\bar r} p \,,
\ee
and
\ba
 &&\left( e^{\bar B + d \overline W} \partial_{\bar \rho}
   \alpha^a \right) \left[ \frac{\partial}{\partial \alpha^a}
    \left( e^{B_f+dW_f} \partial_\rho W_f \Bigr]
    \right) \right]_{\rho = \overline \rho} \\
 && \qquad \qquad \qquad \qquad
 = \left\{ \left[ \bar r \partial_{\bar r} p \left(
 \frac{\partial}{\partial p} \right)
 + \bar r \partial_{\bar r} l_{\ssW f} \left(
 \frac{\partial}{\partial l_{\ssW f}} \right) \right]
 \left[ - \frac{2p}{d-1} \left( \frac{r^{2p}}{r^{2p}
 + l_{\ssW f}^{2p}} \right) \right] \right\}_{r=\bar r} \,.
 \nn
\ea

Rather than using these expressions, however, it is much more
convenient to use (\ref{RGU2'}) to determine $\renctwoTensn$, and
then directly use the constraint, eq.~\pref{braneconstraintren},
to find $\renctwoPotl$. We now illustrate how this works in more
detail by considering a simple example.

\subsubsection*{An example}

To better understand the RG equation's implications, we
follow\footnote{For notational simplicity we drop the bars over
$\phi$ in this section. Our conventions make our couplings
$\lambda_{2n}$ larger than those of ref.~\cite{GW} by a factor of
$2\pi$.} \cite{GW} and expand the $\phi$-dependence of the
codimension-2 tension in a complete basis,
 \be
 \redtwoTensn(\phi) = \sum_{n=0}^{\infty}
 \,\lambda_{2n} \,
 \frac{\phi^{2n}}{(2 n)!} \,,
 \label{gwU2}
\ee
where the constants $\lambda_{2n}$ are effective coupling
constants that control the coupling of the bulk scalar to the
brane, that are $\phi$-independent by definition. Since $\phi =
\phi(\bar r)$ depends explicitly on $\bar r$ (through the bulk
scalar profile) while $\redtwoTensn '$ does not, the renormalized
couplings $\lambda_{2n}$ must depend implicitly on $\bar r$. Our
goal is to use the RG equations to extract this dependence
explicitly.

To this end we insert the form (\ref{gwU2}) in equation
(\ref{RGU2'}), obtaining
\ba \label{RGDE1}
    0 = {\cal D} \redtwoTensn '
    &=& \sum_{n=1}^{\infty} \left[ {\cal D} \lambda_{2n}
    \, \frac{\phi^{2n-1}}{(2n-1)!} +
    \lambda_{2n} \frac{\phi^{2n-2}}{(2n-2)!}
    \, {\cal D} \phi \right] \nn\\
    &=& \sum_{n=1}^{\infty} \left[
    {\cal D} \lambda_{2n} \, \frac{\phi^{2n-1}}{(2n-1)!}
    + \lambda_{2n} \frac{\phi^{2n-2}}{(2n-2)!}
    \left( \sum_{p=1}^{\infty} \,\lambda_{2p}
    \frac{\phi^{2p-1}}{(2p-1)!} \right) \right]
  \nn \\
  &=& \sum_{n=1}^{\infty}\,c_{2n}\,\frac{\phi^{2n-1}}{(2n-1)!}
  \label{rgescal}
\ea
where to pass from the first to the second line we use the dilaton
junction condition (\ref{dilatonJCbar}), ${\cal D} \phi =
{\redtwoTensn}'$, and the last line re-orders the sums to define
\be \label{defc2n}
 c_{2n} \equiv {\cal D} \lambda_{2n} +
 \sum_{k=1}^n \, \left( {\begin{array}{*{20}c} 2n-1
 \\ 2k -1 \\ \end{array}} \right)
 \, \lambda_{2k}\,\lambda_{2n-2k+2} \,.
\ee
Crucially, the condition ${\cal D} {\redtwoTensn}' = 0$ applies as
an identity for all values of the integration constants
characterizing the bulk fields --- like $\phi_b$, $\xi $, $R$ {\it
etc.} --- provided these are held fixed when $\bar r$ is varied.
In particular, although the couplings $\lambda_{2n}$ can also
depend on some of these parameters, they contain enough freedom to
vary $\phi$ with the $\lambda_{2n}$'s held fixed. This implies
that eq.~\pref{RGDE1} holds as an identity for all $\phi$, and so
all the quantities $c_{2n}$ must separately vanish. In this way,
we obtain the following renormalization group equations for the
couplings $\lambda_{2n}$, for $n\ge1$,
\be \label{RGDE2}
 {\cal D} \lambda_{2n} = \xi \hat r \,
 \frac{\partial \lambda_{2n}}{\partial \hat r} = -
 \sum_{k=1}^n \, \left( {\begin{array}{*{20}c} 2n-1
 \\ 2k -1 \\ \end{array}} \right)
 \, \lambda_{2k}\,\lambda_{2n-2k+2} \,,
\ee
where $\hat r = \bar r - l$. Restricting to flat geometries having
conical singularities, and keeping in mind that $\xi = \alpha$ for
such geometries, these RG equations become those obtained in
\cite{GW} by means of graphical methods. Our formalism, then,
easily captures the RG evolution of the brane couplings, without
the need of going through the perturbative calculations used in
the previous literature.

Similar steps can be used to derive RG equations for the analogous
couplings in $\redtwoPotl$,
\be
 \redtwoPotl(\phi) = \sum_{n=0}^{\infty}
 \,\gamma_{2n} \,
 \frac{\phi^{2n}}{(2 n)!} \,,
 \label{gwT2}
\ee
but a simpler procedure to find the $\gamma_{2n}$'s is to directly
use the curvature constraint to relate them to the
$\lambda_{2n}$'s. Working to leading order in $\kappa^2$ implies
$\redtwoPotl \simeq \frac12 \left( {\redtwoTensn}' \right)^2$ and
so
\be
 \gamma_{2n} \simeq \frac12 \sum_{k=1}^n
 \left( {\begin{array}{*{20}c}
 2n \\ 2k - 1 \\ \end{array}} \right)
 \lambda_{2k} \lambda_{2n-2k+2} \,.
\ee

Notice that none of these expressions provide the renormalization
group equation for the coupling $\lambda_0$. To obtain this we
turn to the third RG equation, (\ref{RGU2}). Direct application of
${\cal D}$ to eq.~\pref{gwU2} implies
\be
 {\cal D} \, \redtwoTensn
 = {\cal D} \lambda_{0} + \sum_{n=1}^{\infty} \,
 \left[ {\cal D} \lambda_{2n} + \sum_{k=1}^n\,
 \left( {\begin{array}{*{20}c} 2n \\ 2k -1 \\
 \end{array}} \right) \, \lambda_{2k}\,\lambda_{2n-2k+2}
 \right] \,\frac{\phi^{2n}}{(2n)!} \,.
\ee
Evaluating ${\cal D} \lambda_{2n}$ with eq.~\pref{RGDE2}, and
using the identity
\be
 \left( {\begin{array}{*{20}c} 2n \\ 2k -1 \\
 \end{array}} \right) = \left( {\begin{array}{*{20}c} 2n-1 \\ 2k
 -1 \\ \end{array}} \right) + \left( {\begin{array}{*{20}c} 2n-1 \\
 2k -2 \\ \end{array}} \right) \,,
\ee
we find
\be
  0 = {\cal D} \lambda_{0} + \sum_{n=1}^{\infty}\,
  \sum_{k=1}^n\, \left( {\begin{array}{*{20}c} 2n-1 \\
  2k -2 \\ \end{array}} \right) \,
  \lambda_{2k}\,\lambda_{2n-2k+2}
  \, \left[ \frac{\phi^{2n}}{(2n)!} \right] \,.
\ee
This expression simplifies with the following manipulations:
\ba
 \sum_{n=1}^{\infty} \, \sum_{k=1}^n\, \left(
 {\begin{array}{*{20}c} 2n-1 \\ 2k -2 \\ \end{array}}
 \right) \, \lambda_{2k}\,\lambda_{2n-2k+2}
 \, \left[ \frac{\phi^{2n}}{(2n)!} \right] &=& \frac12\, \left(
 \sum_{n=1}^{\infty} \lambda_{2n}\,\frac{\phi^{2n-1}}{(2n-1)!}
 \right)^2 \\
 &=& \frac12 \left( {\redtwoTensn}' \right)^2
 = \frac{{\xi}^2\,p_{\phi}^2}{2} \,,
\ea
where the last equality uses the dilaton junction condition,
(\ref{dilatonJCbar}).

The evolution equation for $\lambda_0$ then becomes
\be
 {\cal D} \lambda_{0} + \frac{{\xi}^2 \,p_{\phi}^2}{2} =
 \left\{ \left( \bar r \partial_{\bar r}
 p \, \frac{\partial}{\partial p} + \bar r \partial_{\bar r}
 l_{\ssW f} \, \frac{\partial}{\partial l_{\ssW f}} \right)
 \left[ p \left( \frac{l_{\ssW f}^{2p} - r^{2p}
 }{l_{\ssW f}^{2p} + r^{2p}}
 \right) \right] \right\}_{r = \bar r}
 \simeq \bar r \partial_{\bar r} p \,,
\ee
where the approximate equality neglects $\bar r^2 R \propto (\bar
r/l_{\ssW f})^{2p}$ ({\it c.f.} eq.~\pref{Rfsoln}). Using $r_b/l =
\xi/(\xi-1)$ we find that neglect of $\bar r^2 R$ allows
eqs.~\pref{extrensoln2} and \pref{psoln} to simplify to
\ba \label{pformula}
 p &\simeq& e^{d \overline{W}} = \left[ \left(\xi-1\right)
 \,\left( \frac{\bar r - l}{l}
 \right) \right]^{d(\Omega - p_\ssB)/(d-1)} \nn\\
 &\simeq& 1 + \frac{p_\phi^2}{2 p_\ssB}
 \ln \left[ (\xi - 1) \left( \frac{\bar r - l}{l}
 \right) \right] + \cdots \,,
\ea
which uses $d(\Omega - p_\ssB)/(d-1) \simeq p_\phi^2/(2p_\ssB)
\simeq \kappa^2 \ctwoPotl/(2\pi \xi) \ll 1$ in the weak-gravity
limit, using eq.~\pref{BulkFieldAsym3app}. Notice that $p \to 1$
as $\bar r \to r_b$, and because $\Omega \ge p_\ssB$ (with $\Omega
= p_\ssB$ only when $p_\phi = 0$) $p$ diverges as $l \to 0$.

Writing $\bar r \partial_{\bar r} p = {\cal D} p$ we see that
${\cal D}(\lambda_0 - p) + \frac12 \, \xi^2 p_\phi^2 \simeq 0$,
which admits the simple solution
\be
 \lambda_0 = \lambda_{0b} -1+ \left[ \left(\xi-1\right)
 \,\left( \frac{\bar r - l}{l}
 \right) \right]^{d(\Omega - p_\ssB)/(d-1)}
 -\frac{\xi\, p^2_{\phi}}{2} \,
 \ln \left[ (\xi - 1) \left( \frac{\bar r - l}{l}
 \right) \right] \,.
\ee
Using the weak-gravity limit --- {\it i.e.} the second line of
eq.~\pref{pformula} and the leading approximation, $\xi
p_B\,\simeq\,1$, to eq.~(\ref{BulkFieldAsymptotics2}) --- allows
this solution to be rewritten
\be
 \lambda_0 \simeq \lambda_{0b}+
 \frac{\xi\, p^2_{\phi}}{2\,\xi\,p_b} \,
 \ln \left[ (\xi - 1) \left( \frac{\bar r - l}{l}
 \right) \right] -
 \frac{\xi\, p^2_{\phi}}{2} \,
 \ln \left[ (\xi - 1) \left( \frac{\bar r - l}{l}
 \right) \right] \simeq \lambda_{0b} \,,
\ee
which shows that $\lambda_0$ does not renormalize up to ${\cal
O}(\kappa^2)$. This holds in particular for the special case of
pure tension branes, for which $p_\phi=0$, and so $\Omega =
p_\ssB$.

In general, we see from this section how to define a complete set
of RG equations for the brane-$\phi$ couplings contained in the
brane action, $\redtwoTensn $, and on-brane potential,
$\redtwoPotl $, generalizing earlier discussions to more general
bulk configurations.

\section{Conclusions}

In summary, this paper uses the example of a scalar-tensor theory
in $D = d+2$ dimensions to examine the detailed connection between
the properties of a $d$-dimensional, codimension-2 brane and the
bulk fields which it supports. The brane in question can be
fundamental ({\it e.g.} a D-brane in string theory) or a
low-energy artifact (like a string defect in a gauge theory),
provided the length scale associated with any brane structure is
much smaller than the scales associated with the fields to which
it gives rise.

Our strategy for identifying this connection is to temporarily
adopt a codimension-1 crutch. That is, we first regulate the
codimension-2 brane as a very small codimension-1 object. The
codimension-2 action is connected by dimensional reduction to the
codimension-1 one, which is in turn related to the bulk properties
by standard junction conditions. Once the connection between bulk
and codimension-2 properties is made we kick the crutch away,
confident that its details are not important for the purposes of
describing only the leading low-energy behaviour.

We find the following results
\begin{itemize}
\item In codimension two the bulk fields generically diverge as
one approaches the brane sources, and this divergence is not
restricted to a purely conical defect. Typically the appearance of
curvature singularities at the brane position signals a nontrivial
coupling between the brane and the bulk scalar.
\item There are two quantities that characterize the properties of
codimension-2 branes at low energies: the effective brane tension,
$\ctwoTensn(\phi)$, and the brane contribution to the effective
on-brane scalar potential, $\ctwoPotl(\phi)$. From the point of
view of the codimension-1 regulating brane these two quantities
respectively correspond to the on-brane and `angular'
stress-energies, $T_{\mu\nu}$ and $T_{\theta\theta}$,
dimensionally reduced in the angular direction.
\item The codimension-2 brane tension sources the bulk scalar
field in the way one would naively expect for a $\delta$-function
source, with its derivative, ${\ctwoTensn}'$, controlling the
appropriately-defined near-brane radial derivative of the scalar
field, $\partial_r \phi$. The on-brane potential, $\ctwoPotl$,
similarly contributes in the usual way to the low-energy dynamics
of any light KK zero modes, including the curvature of the
low-energy metric through the low-energy Einstein equations.
\item The field equation impose a general constraint relating
these quantities to the on-brane curvature, that provides the
generalization of the codimension-1 brane modification to the
Friedmann equation. We argue that for codimension-2 branes its
proper interpretation within a low-energy framework is as a
constraint that relates $\ctwoPotl$ to $\ctwoTensn$: $4 \pi
\ctwoPotl \simeq \kappa^2 \left( {\ctwoTensn}' \right)^2$. This
relation shows that any dynamics that causes $\phi$ to make
$\ctwoPotl$ small (and so minimize the brane's contribution to the
low energy on-brane curvature), also minimizes its coupling to the
codimension-2 brane tension.
\end{itemize}

All of these results are prerequisites for the exploration of the
utility of codimension-2 branes for addressing low-energy problems
in particle physics and cosmology, a direction of research we hope
this paper encourages.

\section*{Acknowledgements}

We wish to thank Fernando Quevedo and Andrew Tolley for many
helpful comments and suggestions. This research has been supported
in part by funds from the Natural Sciences and Engineering
Research Council (NSERC) of Canada. CB also acknowledges support
from CERN, the Killam Foundation, and McMaster University. CdR is
partly funded by an Ontario Ministry of Research and Information
(MRI) postdoctoral fellowship. GT thanks Perimeter Institute for
kind hospitality during the beginning of this work. Research at
the Perimeter Institute is supported in part by the Government of
Canada through NSERC and by the Province of Ontario through MRI.

\appendix

\section{General Axial Solutions to the Field Equations
when $V=0$} \label{App:SolvingFieldEq}

This appendix provides details of how the bulk field equations in
$D$ spacetime dimensions are integrated for geometries in the case
$V=0$, subject to the symmetry ansatz of axial symmetry in the
transverse two dimensions spanned by $(\rho,\theta)$ and maximal
symmetry in the $d = D-2$ dimensions spanned by $x^\mu$.

\subsection*{Bulk Equations}

The field equations to be solved when $V=0$ are
\ba \label{app:bulkrhoeqns}
    \phi'' + \Bigl\{ d\,W' + B' \Bigr\} \phi' &=& 0
    \quad \hbox{($\phi$)}\nn\\
    \frac{\cR}{d} +  W'' + d\, (W')^2 + W' B'
    &=& 0 \quad \hbox{($\mu\nu$)} \nn\\
    B'' + (B')^2 + d\, W' B'  &=& 0
    \quad \hbox{($\theta\theta$)} \nn\\
   d\, \Bigl\{ W'' + (W')^2 \Bigr\} + B'' + (B')^2 +
    (\phi')^2  &=& 0
    \quad \hbox{($\rho\rho$)} \,.
\ea
where we use the conventions defined in the main text. The general
solution to these may be written down in closed form as follows. A
first integral of the dilaton and ($\theta\theta$) Einstein
equations can be done by inspection to give
\be \label{app:phi'B'}
    e^{B +d\, W} \phi' = \hat p_\phi
    \qquad \hbox{and} \qquad
    e^{B + d\, W} B' = \hat p_\ssB \,,
\ee
for $\hat p_\phi$ and $\hat p_\ssB$ arbitrary constants. These may
both be integrated a second time by conveniently redefining a new
radial coordinate $r$ as
\be
    \frac{\d r}{r} := \xi \, e^{-B-d W} \d \rho \,,
\ee
in terms of which $e^{B+d W}\partial_\rho = \xi \, r
\,\partial_r$. This leads to the solutions
\be \label{app:phiBsolns}
    e^\phi = e^{\phi_0} \left( \frac{r}{l} \right)^{p_\phi}
    \qquad \hbox{and} \qquad
    e^B = l \left( \frac{r}{l} \right)^{p_\ssB} \,,
\ee
with $p_\phi = \hat p_\phi/\xi$ and $p_\ssB = \hat p_\ssB/ \xi$,
and new integration constants $\phi_0$ and $l$.

Similarly the combination $(\rho\rho) - (\theta\theta)$ of
Einstein equations gives
\be \label{app:rho-theta}
    d \Bigl\{W'' + (W')^2 - W' B' \Bigr\}
    + (\phi')^2  = 0 \,.
\ee
Multiplying this through by $e^{2(B +d W)}$, and changing
variables from $\rho$ to $r$, using
\ba
    \xi^2 (r \, \partial_r)^2 W
    &=& e^{B+d W} \Bigl( e^{B+d W} W'
    \Bigr)' \nn\\
    &=&  e^{2[B+d W]} W'' + \xi^2 \Bigl[
    (r \, \partial_r B)(r \,
    \partial_r W) + d  (r \, \partial_r W)^2 \Bigr]\,,
\ea
then gives
\be
    \ddot W - (d-1)(\dot W)^2 - 2\dot B \dot W
    + \frac{(\dot \phi)^2}{d}
    = 0 \,.
\ee
Here over-dots denoting differentiation with respect to $\ln r$.
Using eqs.~\pref{app:phiBsolns} for $\phi$ and $B$, leads to a
differential equation involving only $W$
\be
    \ddot W - 2 p_\ssB \dot W - (d-1)(\dot W)^2
    + \frac{p_\phi^2}{d}
    = 0 \,.
\ee
This equation can be regarded as a first-order equation for $\dot
W$, and so may be directly integrated twice, leading to the
following general solution
\ba
    (d-1) W &=& (d-1) W_0 - p_\ssB \ln \left(\frac{r}{l} \right)
    - \ln \cosh X \nn\\
    \hbox{with}\quad
    X &=& \Omega \, \ln\left( \frac{r}{l_\ssW} \right) \nn\\
    \hbox{and} \quad
    \Omega &=& \sqrt{ p_\ssB^2 + \left( \frac{d-1}{d}
    \right) p_\phi^2} \,,
\ea
where $W_0$ and $l_\ssW$ are the two new integration constants. We
find a total of six integration constants --- $\phi_0$, $W_0$,
$l$, $l_\ssW$, $p_\phi$ and $p_\ssB$ --- of which one ($W_0$) can
be changed simply by re-scaling $x^\mu$.

Finally, the on-brane induced curvature scalar, $R$, may be
obtained using the ($\mu\nu$) Einstein equation, which states
\be
    \xi^2 \ddot W
    = e^{B + d W} \Bigl(  e^{B + d W} W'  \Bigr)'
    = - \frac{e^{2[B + d W]} \cR}{d} =
    - \frac{e^{2[B + (d-1)W]} R}{d}  \,.
\ee
Using the explicit form just found for the solution,
\be
    e^{B+(d-1)W} = \frac{l \,e^{(d-1)W_0}}{\cosh X}
    = \frac{2\,l\, e^{(d-1)W_0}}{ (r/\ell_\ssW)^\Omega
    + (\ell_\ssW/r)^\Omega }\,,
\ee
as well as
\be\label{exprfW}
    \ddot W = - \left( \frac{\Omega^2}{d-1} \right)
    \frac{1}{\cosh^2 X} \,,
\ee
we find in this way
\be\label{exprfR}
    R = - d \, \xi^2 \,\ddot W \, e^{-2[B+(d-1)W]}
    = \left[ \frac{d \, \xi^2 \,\Omega^2}{(d-1) \, l^2}
    \right] \, e^{-2(d-1)W_0} \,.
    \ee

Given these explicit functions for $B$ and $W$, we may compute the
relation between $r$ and proper distance, $\rho$, by integrating
\ba
    \xi\, \d\rho &=& e^{B + d W} \, \frac{\d r}{r} \nn\\
    &=& l \, e^{d W_0} \left( \frac{l}{r}
    \right)^{p_\ssB/(d-1)} \left[ \frac{2}{(r/l_\ssW)^\Omega
    + (l_\ssW/r)^\Omega} \right]^{d/(d-1)}
    \frac{\d r}{r}\,,
\ea
which implies $\rho \propto r^{[- p_\ssB -d \Omega ]/(d-1)}$ in
the limit $r \gg l_\ssW$ while $\rho \propto r^{[ -p_\ssB + d
\Omega]/(d-1)}$ when $r \ll l_\ssW$.

\subsubsection*{The flat limit}

The special case of the flat limit, $R\to 0$, can be seen to
correspond to the choice $W_0 \to \infty$ and $l_\ssW \to \infty$,
with the ratio $e^{(d-1)W_0}/l_\ssW^\Omega \equiv \frac12 \,
e^{(d-1)w_0} l^{-\Omega}$ fixed, since in this case the above
expressions for $\phi$ and $B$ are unchanged, while
\be
    e^{(d-1)W} = \frac{2e^{(d-1)W_0}}{ (r/l_\ssW)^\Omega
    + (l_\ssW/r)^\Omega } \left( \frac{l}{r} \right)^{p_\ssB}
    \to e^{(d-1)w_0}
    \left( \frac{r}{l} \right)^{\Omega-p_\ssB} \,,
\ee
so the resulting solution is
\be \label{app:powerlaws}
    e^\phi = e^{\phi_0} \left( \frac{r}{l}
    \right)^{p_\phi} \,, \quad
    e^B = l \left( \frac{r}{l}
    \right)^{p_\ssB}
    \quad \hbox{and}\quad
    e^{(d-1)W} = e^{(d-1)w_0} \left( \frac{r}{l}
    \right)^{\Omega - p_\ssB} \,.
\ee

It is useful to re-express these solutions in terms of the proper
distance
\be
    \xi\, \rho = \left[ \frac{l \, (d-1)}{-p_\ssB
    + d \Omega} \right] \, e^{d w_0} \left( \frac{r}{l}
    \right)^{[-p_\ssB + d \Omega]/(d-1)} \,,
\ee
giving the convenient form
\be \label{app:powerlaws2}
    e^\phi = e^{\phi_0} \left( \frac{\rho}{\ell}
    \right)^{\gamma} \,, \quad
    e^B = \ell \left( \frac{\rho}{\ell}
    \right)^{\beta}
    \quad \hbox{and}\quad
    e^{W} = e^{w_0} \left( \frac{\rho}{\ell}
    \right)^{\omega} \,,
\ee
where $\ell$ is a constant in principle calculable in terms of
$l$, $\xi$, $p_\ssB$ {\it etc.}, and the powers satisfy
\be \label{app:kasnerconditions}
    d \, \omega^2 + \beta^2 + \gamma^2
    = d \,\omega + \beta = 1\,.
\ee
In terms of these the derivatives appearing in the jump conditions
are
\be
    \partial_\rho \phi = \frac{\gamma}{\rho} \,,
    \quad
    \partial_\rho B = \frac{\beta}{\rho} \quad
    \hbox{and} \quad
    \partial_\rho W = \frac{\omega}{\rho}  \,.
\ee

\section{Derivation of the Codimension-1 Matching Conditions}
\label{App:jumpconds}

Here derive the matching conditions in detail

\subsection*{Gauss-Codazzi Equations}

Consider a $D$-dimensional geometry which in some region is
foliated into a series of surfaces, $\Sigma$. The Gauss-Codazzi
equations express the Riemann tensor of the full space in terms of
the intrinsic and extrinsic curvatures on these surfaces. To
derive these expressions, choose coordinates in the region of
interest so that the surfaces are surfaces of constant coordinate,
$\rho$, and for which the metric is
\be
    \d s^2 = \d \rho^2 + \hg_{mn} \, \d x^m \, \d x^n \,.
\ee
$\rho$ clearly measures the proper distance between the surfaces.
In these coordinates $\hg_{mn} = \hg_{mn}(\rho,x)$ defines the
intrinsic geometry on the surfaces $\Sigma$. The intrinsic
curvature tensor, $\hRmu{m}{nrs}$, is defined in the usual way
from the Christoffel symbols, $\hG^m_{nr} = \frac12 \, \hg^{ms}
(\partial_n \hg_{rs} + \partial_r \hg_{ns} -
\partial_s \hg_{nr})$, by\footnote{These follow Weinberg's
curvature conventions, and so only differ from MTW's by an overall
sign.}
\be
    \hRmu{m}{nrs} = \partial_s \hG^m_{nr} + \hG^m_{sq}
    \hG^q_{nr} - (r \leftrightarrow s) \,.
\ee

The extrinsic curvature, $K_{mn}$, is similarly defined in terms
of the unit normal, $N_\ssM \d x^\ssM = \exd \rho$, of $\Sigma$,
by
\be
    K_{\ssM\ssN} = {P_\ssM}^\ssP {P_\ssN}^\ssR \nabla_\ssP N_\ssR
\ee
where ${P_\ssM}^\ssN = \delta_\ssM^\ssN - N_\ssM N^\ssN$ is the
projector onto $\Sigma$, and so in the given coordinates we have
\be
    K_{mn} = \partial_m N_n - \Gamma^\ssM_{mn} N_\ssM
    = - \Gamma^\rho_{mn} = \frac12 \, \partial_\rho \hg_{mn} \,,
\ee
which uses the following expressions for the Christoffel symbols
for the full, bulk metric:
\be
    \Gamma^m_{nr} = \hG^m_{nr} \,, \qquad
    \Gamma^\rho_{mn} = - \frac12 \, \partial_\rho \hg_{mn}
    = - K_{mn} \quad \hbox{and} \quad
    \Gamma^m_{\rho n} = + \frac12 \hg^{mr} \partial_\rho \hg_{rn}
    = {K^m}_n \,,
\ee
and $\Gamma^\rho_{\rho m} = \Gamma^m_{\rho\rho} =
\Gamma^\rho_{\rho\rho} = 0$.

Direct use of the definitions gives the components of the full
bulk Riemann tensor as
\ba
    \Rml{mnrs} &=& \hRml{mnrs} - K_{ms} K_{nr} + K_{mr} K_{ns}
    \nn\\
    \Rml{\rho mnr} &=& \hat\nabla_n K_{mr} - \hat\nabla_r K_{mn}
    \nn\\
    \Rml{\rho m \rho n} &=& \partial_\rho K_{mn} - K_{ms} {K^s}_n
    = \nabla_\rho K_{mn} + K_{ms} {K^s}_n \,,
\ea
with any component not related to these by the symmetries of the
Riemann tensor vanishing. The last line defines the quantity
\ba
    \nabla_\rho K_{mn} &=& \partial_\rho K_{mn} - \Gamma^s_{\rho m}
    K_{sn} - \Gamma^s_{\rho n} K_{sm} \nn\\
    &=& \partial_\rho K_{mn} - 2 K_{ms} {K^s}_n \,.
\ea

The components of the Ricci tensor, $\Rc{\ssM\ssN} =
\Rmu{\ssP}{\ssM\ssP\ssN}$, then become
\ba
    \Rc{mn} &=& \hRc{mn} + \partial_\rho K_{mn}
    - 2 K_{ms} {K^s}_n + K \, K_{mn} \nn\\
    &=& \hRc{mn} + \nabla_\rho K_{mn} + K \, K_{mn} \nn\\
    \Rc{\rho m} &=& \partial_m K - \hat\nabla^n K_{nm} \nn\\
    \Rc{\rho\rho} &=& \partial_\rho K + K_{mn} K^{mn} \,,
\ea
where $K = \hg^{mn} K_{mn} = {K^m}_m$. The scalar curvature is
similarly given by
\be \label{app:RicciScalarDecomp}
    R = \hat R + 2 \,\partial_\rho K + K_{mn} K^{mn} + K^2 \,.
\ee

Notice that
\be
    \partial_\rho \Bigl( \sqrt{-g} \Bigr)
    = \frac12 \sqrt{-g} \; \hg^{mn}
    \partial_\rho \hg_{mn} = \sqrt{-g} \; K \,,
\ee
which also implies the following identity
\be
    \partial_\rho \Bigl( \sqrt{-g} \; F \Bigr) = \sqrt{-g}
    \; \Bigl( \partial_\rho F + K F \Bigr) \,,
\ee
when $F$ is any scalar quantity. Applied to the Einstein-Hilbert
action this implies
\ba \label{app:EHDeriv}
    \sqrt{-g} \; R &=& \sqrt{-\hg} \; \Bigl( \hR + 2\,
    \partial_\rho K + K_{mn} K^{mn} + K^2 \Bigr) \nn\\
    &=& \partial_\rho \Bigl(2\, \sqrt{-g} \; K \Bigr) +
    \sqrt{-g} \; \Bigl( \hR + K_{mn} K^{mn} - K^2 \Bigr) \,.
\ea

Finally, the components of the full Einstein tensor, $G_{\ssM\ssN}
= R_{\ssM\ssN} - \frac12 \, R \, g_{\ssM\ssN}$, are given by
\ba  \label{app:EinsteinDecomp}
    G_{mn} &=& \hat{G}_{mn} + \nabla_\rho \Bigl( K_{mn}
    - K \, \hg_{mn} \Bigr) + K K_{mn} - \frac12 \Bigl(
    K_{rs} K^{rs} + K^2 \Bigr) \nn\\
    G_{\rho m} &=& \partial_m K - \hat\nabla^n K_{nm} \nn\\
    G_{\rho\rho} &=& \frac12 \Bigl( K_{mn} K^{mn} - K^2 - \hat R
    \Bigr) \,.
\ea
Notice in particular how the second derivatives of the form
$\partial_\rho^2 \hg_{mn}$ drop out of the expressions for
$G_{\rho m}$ and $G_{\rho\rho}$, making these constraints for the
purposes of integrating the equations in the $\rho$ direction from
given `intial' data at $\rho = \rho_0$.

\subsection*{Actions}

We next turn to how these expressions are to be used to find the
bulk solutions given the properties of a codimension-1 brane. This
starts with the specification of a bulk and brane action, which
can be specified in one of two equivalent ways. First, one can
work within a bulk region, $M$, without boundaries (say), with the
brane contributions explicitly inserted as delta function sources.
That is, write $S = \int_M \d^Dx \; \L$, with
\be
    \L = \L_\ssB(\phi,A_\ssM,g_{\ssM\ssN}) +
    \delta(\rho - \rho_b) L_b(\phi,A_\ssM,g_{\ssM\ssN}) \,,
\ee
so $S = S_B + S_b$, with $S_B = \int_M \d^Dx \; \L_B$ and $S_b =
\int_\Sigma \d^{D-1}x \; \L_b$. In this case the delta-function
source in the equations of motion gives rise to step
discontinuities in the $\rho$-derivatives of the bulk fields, as
can be schematically inferred by integrating the field equations
over a narrow region $\rho_b - \epsilon < \rho < \rho_b +
\epsilon$ in a particular coordinate system (like the one used
above).

Alternatively, we can divide $M$ into the two parts, $M_\pm$,
lying on either side of $\Sigma$, with $M_+$ defining the region
$\rho > \rho_b$ and $M_-$ denoting $\rho < \rho_b$. In this case
we define the bulk action in the regions $M_\pm$ including their
boundaries at $\rho = \rho_b$, and define the brane action only at
$\rho = \rho_b$. In either case the goal is to identify how the
brane action governs the discontinuities of the bulk fields at the
brane position.

\subsubsection*{The Gibbons-Hawking Action}

Because the second approach explicitly involves boundaries it is
necessary to be careful about boundary contributions to actions in
general, and to the gravitational action in particular. The usual
gravitational action is the sum of a bulk (Einstein-Hilbert) and a
boundary (Gibbons-Hawking) part, $S_g = S_{EH} + S_{GH}$, where
\be
    S_{EH}(M) = - \frac{1}{2\kappa^2} \int_M \d^Dx \;
    \sqrt{-g} \; R \,,
\ee
with $\kappa^2 = 8\pi G$ related to the $D$-dimensional Newton
constant.

But eq.~\pref{app:RicciScalarDecomp} shows that this action
contains terms like $\partial_\rho^2 \, \hg_{mn}$, and so on
variation contains boundary terms of the form $\partial_\rho
\delta \hg_{mn}$. Since these derivatives can be varied
independently from $\delta \hg_{mn}$ on the boundary, the result
is an over-constrained problem with excessively constrained
boundary information. This fact does not normally cause problems
when formulating solutions to Einstein's equations without
boundaries, because eq.~\pref{app:EHDeriv} shows that these terms
enter in a total derivative. When boundaries are present, the
second derivative terms must be explicitly subtracted by
supplementing the action by the appropriate boundary term:
\ba
    S_{GH} &=& +\frac{1}{\kappa^2} \int_{\partial M} \d^{D-1}x \;
    \sqrt{-\hg} \; K \nn\\
    &=& \frac{1}{\kappa^2} \int_{\Sigma(\rho_{\rm max})}
    \d^{D-1}x \; \sqrt{-\hg} \; K
    - \frac{1}{\kappa^2} \int_{\Sigma(\rho_{\rm min})}
    \d^{D-1}x \; \sqrt{-\hg} \; K \,.
\ea
With this choice the total gravitational action decomposes as
follows
\be \label{app:TotGravAction}
    S_g = S_{EH} + S_{GH}
    = - \frac{1}{2\kappa^2} \int_M \d^Dx \; \sqrt{-g} \;
    \Bigl( \hat{R} + K_{mn} K^{mn} - K^2 \Bigr) \,.
\ee

\subsection*{Jump Conditions}

The next step is to derive the coupling between brane and bulk in
the field equations.

\subsubsection*{Israel Junction Condition}

Once the surface action has been added to the gravitational
kinetic term, it is possible to keep track of how its variation
depends on the variation of the metric on the boundaries. Keeping
track only of the boundary terms in the variation of
eq.~\pref{app:TotGravAction} leads to
\be
    \delta S_g = - \frac{1}{2\kappa^2} \int_{\partial M}
    \d^{D-1}x \; \sqrt{-\hg} \; \Bigl( K^{mn} - K \hg^{mn}
    \Bigr) \delta \hg_{mn} + \cdots \,,
\ee

For a brane spanning the surface $\Sigma$ at $\rho = \rho_b$ lying
between the two regions $M_\pm$ the total contribution to the
equations of motion coming from variations of the boundary metric
then is
\be \label{app:IsraelJC1}
    \frac{1}{2\kappa^2} \Bigl[ \sqrt{-\hg} \; \Bigl(
    K^{mn} - K \hg^{mn} \Bigr) \Bigr]_{\rho_b} +
    \frac{\delta S_b}{\delta \hg_{mn}} = 0 \,,
\ee
where the notation $[ F ]_{\rho_b}$ for a bulk quantity denotes
the jump
\be
    \Bigl[ F \Bigr]_{\rho_b} = \lim_{\epsilon \to 0} \Bigl[
    F(\rho_b + \epsilon) - F(\rho_b - \epsilon) \Bigr] \,.
\ee
Denoting the stress energy for the bulk and brane by
\be
    T^{\ssM\ssN} = \frac{2}{\sqrt{-g}} \, \frac{\delta S_B}{\delta
    g_{\ssM\ssN}} \qquad \hbox{and} \qquad
    t^{mn} = \frac{2}{\sqrt{-\hg}} \, \frac{\delta S_b}{\delta
    \hg_{mn}} \,,
\ee
the Israel jump condition becomes
\be \label{app:IsraelJC2}
    \Bigl[ K_{mn} - K \hg_{mn} \Bigr]_{\rho_b} + \kappa^2 \,
    t_{mn} = 0 \,.
\ee

Notice that this condition could equivalently be derived in the
delta-function formulation of the action, by isolating the
delta-function contribution to the LHS and RHS of the $(mn)$
Einstein equation:
\ba
    0 &=&\lim_{\epsilon \to 0} \int_{\rho_b - \epsilon}^{\rho_b +
    \epsilon} \d\rho \; \Bigl\{ \Bigl[ G_{mn} +
    \kappa^2 \, T_{mn} \Bigr] + \delta(\rho - \rho_b)
    \; \kappa^2 \, t_{mn} \Bigr\}  \nn\\
    &=& \Bigl[ K_{mn} - K \, \hg_{mn} \Bigr]_{\rho_b} + \kappa^2
    \, t_{mn} \,,
\ea
because the step discontinuity in $K_{mn} \propto \partial_\rho
\hg_{mn}$ across the brane implies a delta-function discontinuity
in the contributions of $\partial_\rho K_{mn}$ to the Einstein
tensor (see eq.~\pref{app:EinsteinDecomp}).

Notice also that if the brane represents a physical boundary to
spacetime (rather than being a surface embedded into it), then the
same arguments show that variation of the metric on the boundary
leads to the boundary condition
\be \label{app:MetricBC}
    \pm \frac{1}{2\kappa^2} \sqrt{-\hg} \; \Bigl(
    K^{mn} - K \hg^{mn} \Bigr) +
    \frac{\delta S_b}{\delta \hg_{mn}} = 0 \,,
\ee
where the $+$ sign applies at the boundary at $\rho = \rho_{\rm
max}$ and the $-$ sign applies at $\rho = \rho_{\rm min}$.

\subsubsection*{The Constraints}

To the extent that the brane action does not support any off-brane
components to stress energy, $t_{\rho m} = t_{\rho\rho} = 0$,
there is no discontinuity in the remaining components of the bulk
Einstein equations, which then are
\be
    \partial_m K - \hat\nabla^n K_{nm} + \kappa^2 \, T_{\rho m} =
    0 \,,
\ee
expressing no net energy exchange with the brane, and
\be
    \frac12 \Bigl( K_{mn} K^{mn} - K^2 - \hat{R} \Bigr) + \kappa^2
    \, T_{\rho\rho} = 0 \,,
\ee
which can be solved to give the induced curvature scalar in terms
of the asymptotic forms for the bulk fields, giving
\be
    \hat{R} = K_{mn} K^{mn} - K^2 - 2\, \kappa^2 \, T_{\rho\rho} \,.
\ee

\subsection*{Scalar Jump Condition}

We derive the scalar jump condition in two ways: using an explicit
boundary and thinking of the brane as a delta-function source.

We start with the traditional derivation. If the scalar field
kinetic term has the form
\be \label{app:ScalarKineticTerm}
    S_\phi(M) = - \frac{1}{2\kappa^2} \int_M \d^{D}x \; \sqrt{-g} \,
    g^{\ssM\ssN} \partial_\ssM \phi \, \partial_\ssN \phi \,,
\ee
then these same arguments can be repeated to read off how the
brane action gives rise to derivative discontinuities in $\phi$ at
the brane positions. Since the boundary variation of
eq.~\pref{app:ScalarKineticTerm} is
\be
    \delta S_\phi = - \frac{1}{\kappa^2} \int_{\partial M}
    \d^{D-1}x \; \sqrt{-\hg} \; N^\ssM \partial_\ssM \phi
    \, \delta \phi
    \,,
\ee
where $N_\ssM$ is the outward-pointing normal. Combining the
contributions of regions $M_\pm$ to that of the brane action, and
keeping in mind that $N_\ssM \d x^\ssM = \mp \d\rho$ for the
boundary between these two regions, gives
\be \label{app:ScalarJC1}
    \frac{1}{\kappa^2} \Bigl[ \sqrt{-\hg} \, \partial_\rho \phi
    \Bigr]_{\rho_b} + \frac{\delta S_b}{\delta \phi} = 0 \,.
\ee

Alternatively, let us re-derive the jump condition by regarding
the brane to be a delta-function source to the scalar field
equation. Writing $S_b = \int \d^{D}x \L_b = \int \d^{D-1}x \,
\d\rho \, \L_b \, \delta(\rho - \rho_b)$, we have:
\be
    \sqrt{-\hg} \, \Box \phi + \kappa^2 \, \left(
    \frac{\partial \L_b}{\partial \phi} \right)
    \delta(\rho - \rho_b) = 0\,.
\ee
We next integrate this equation over the disk having radius $\rho
= \rho_b + \epsilon$ and take $\epsilon \to 0^+$. This gives
\be
    \int_0^{\rho_b + \epsilon} \d\rho \, \partial_\rho \Bigl(
    \sqrt{-\hg} \, \hg^{\rho\rho} \partial_\rho \phi \Bigr)
    = \sqrt{-\hg} \, \phi'(\rho_b + \epsilon)
    = - \kappa^2 \, \left(
    \frac{\partial \L_b}{\partial \phi} \right) \,.
\ee
In either case we have the same result:
\be \label{app:ScalarJC2}
    \partial_\rho \phi(\rho \to \rho_b^+)
    = - \frac{\kappa^2}{\sqrt{-\hg}}
    \frac{\delta S_b}{\delta \phi}
    = - \frac{\kappa^2}{\sqrt{-\hg}}
    \frac{\d}{\d \phi} \Bigl( \sqrt{-\hg} \; L_b
    \Bigr)\,,
\ee
where we write $\L_b = \sqrt{-\hg} \; L_b$. For future
applications it is worth noticing that when the brane position
depends on $\phi$ -- {\it i.e.} $\rho_b = \rho_b(\phi)$ -- the
measure $\sqrt{-\hg}$ does as well, and so cannot be pulled out of
the derivative $\d/\d\phi$ to cancel the denominator in the
prefactor.

\section{Matching with Derivative Corrections}
\label{App:MatchingwDerivatives}

For pure tension branes the junction conditions imply that the
discontinuity in the combination $[W'-B']$ vanishes. However this
discontinuity becomes nonzero once derivative terms are included
in the brane action. In this section we compute this correction.

Working to two-derivative order in the brane action we instead
have
\be
    S_b = \int \d^{D-1} x \, \L_1
    = - \int \d^{D-1} x \sqrt{-\hg} \; \left\{ \coneTensn (\phi)
    + \frac12 \, \hg^{mn} \Bigl( X_1(\phi) \, \partial_m \phi \,
    \partial_n \phi + Y_1(\phi) \, \hR_{mn} \Bigr) \right\} \,.
\ee
We imagine working with canonical kinetic terms in the bulk and so
are not free to redefine the metric and scalar to remove the
functions $X_1$ and $Y_1$. Given this action the brane stress
energy becomes
\be
    t^{mn} = - \hg^{mn} \, \left\{ \coneTensn  + \frac12 \, \Bigl(
    X_1 \partial_s \phi \, \partial^s \phi + Y_1 \hR \Bigr)
    + \hat\Box Y_1 \right\} + X_1 \partial^m \phi \, \partial^n \phi
    + Y_1 \hR^{mn} + \hat\nabla^m \hat\nabla^n Y_1 \,. \ee
Using this in the Israel junction condition
\be
    \Bigl[ K_{mn} - K \, \hg_{mn} \Bigr] + \kappa^2 \, t_{mn} = 0
    \,,
\ee
now gives
\ba
    \Bigl[ (D-3) W' + B' \Bigr] \cg_{\mu\nu}
    &=& - \cg_{\mu\nu} \,\kappa^2 \left\{ \coneTensn  + \frac12 \, \Bigl(
    X_1 \partial_s \phi \, \partial^s \phi + Y_1 \hR \Bigr)
    + \hat\Box Y_1 \right\} \nn\\
    &&\qquad\qquad + \kappa^2 \Bigl( X_1 \partial_\mu \phi
    \, \partial_\nu \phi + Y_1 \hR_{\mu\nu} + \hat\nabla_\mu
    \hat\nabla_\nu Y_1 \Bigr) \nn\\
    \Bigl[ (D-2)\, W' \Bigr] g_{\theta\theta}
    &=& - g_{\theta\theta} \,\kappa^2 \left\{ \coneTensn  + \frac12 \, \Bigl(
    X_1 \partial_s \phi \, \partial^s \phi + Y_1 \hR \Bigr)
    + \hat\Box Y_1 \right\} \nn\\
    &&\qquad\qquad + \kappa^2 \Bigl( X_1 \partial_\theta \phi
    \, \partial_\theta \phi + Y_1 \hR_{\theta\theta} + \hat\nabla_\theta
    \hat\nabla_\theta Y_1 \Bigr)
    \,,
\ea
or, equivalently,
\ba
    \Bigl( X_1 \partial_\mu \phi
    \, \partial_\nu \phi + Y_1 \hR_{\mu\nu} + \hat\nabla_\mu
    \hat\nabla_\nu Y_1 \Bigr)
    &=& \frac{1}{D-2} \, \cg_{\mu\nu} \, \cg^{\lambda\sigma}
    \Bigl( X_1 \partial_\lambda \phi
    \, \partial_\sigma \phi + Y_1 \hR_{\lambda\sigma} +
    \hat\nabla_\lambda
    \hat\nabla_\sigma Y_1 \Bigr) \nn\\
    \Bigl[ (D-3) W' + B' \Bigr]
    &=& - \kappa^2 \left\{ \coneTensn  + \frac12 \, \Bigl(
    X_1 \partial_s \phi \, \partial^s \phi + Y_1 \hR \Bigr)
    + \hat\Box Y_1 \right\} \nn\\
    &&\qquad\qquad + \frac{\kappa^2 }{D-2} \,
    \cg^{\mu\nu} \Bigl( X_1 \partial_\mu \phi
    \, \partial_\nu \phi + Y_1 \hR_{\mu\nu} + \hat\nabla_\mu
    \hat\nabla_\nu Y_1 \Bigr) \nn\\
    \Bigl[ (D-2) W' \Bigr]
    &=& - \kappa^2 \left\{ \coneTensn  + \frac12 \, \Bigl(
    X_1 \partial_s \phi \, \partial^s \phi + Y_1 \hR \Bigr)
    + \hat\Box Y_1 \right\} \\
    &&\qquad\qquad + \kappa^2 g^{\theta\theta} \Bigl( X_1 \partial_\theta \phi
    \, \partial_\theta \phi + Y_1 \hR_{\theta\theta} + \hat\nabla_\theta
    \hat\nabla_\theta Y_1 \Bigr)
    \,, \nn
\ea

Now the difference of the last two conditions gives
\be
    \Bigl[ W' - B' \Bigr] =
    \kappa^2 g^{\theta\theta} \Bigl( X_1 \partial_\theta \phi
    \, \partial_\theta \phi + Y_1 \hR_{\theta\theta} + \hat\nabla_\theta
    \hat\nabla_\theta Y_1 \Bigr)
    - \frac{\kappa^2 }{D-2} \,
    \cg^{\mu\nu} \Bigl( X_1 \partial_\mu \phi
    \, \partial_\nu \phi + Y_1 \hR_{\mu\nu} + \hat\nabla_\mu
    \hat\nabla_\nu Y_1 \Bigr) \,.
\ee

Specializing these jump conditions to the case where all
quantities are independent of $\theta$ and the $x^\mu$ directions
are maximally symmetric then reduces them to
\ba
    \Bigl[ (D-3) W' + B' \Bigr]
    &=& - \kappa^2 \left( \coneTensn
    + \frac{D-4}{2 \,(D-2)} \; Y_1 \cR \right) \nn\\
    \Bigl[ (D-2) W' \Bigr]
    &=& - \kappa^2 \left( \coneTensn   + \frac12 \, Y_1 \cR
    \right)
    \,,
\ea
and so
\be
    \Bigl[ W' - B' \Bigr] =
    - \frac{\kappa^2 }{D-2} \,
     Y_1 \cR \,.
\ee

In the general case the scalar jump condition generalizes to
\be
    \Bigl[ \phi' \Bigr] + \frac{\kappa^2}{\sqrt{-\hg}} \,
    \frac{\delta S_b}{\delta \phi} = \Bigl[ \phi' \Bigr]
    - \frac{\kappa^2 }{\sqrt{-\hg}} \,
    \left\{ \sqrt{-\hg} \left[
    \coneTensn  + \frac12 \, \Bigl( X_1 \partial_s \phi
    \, \partial^s \phi + Y_1 \hR \Bigr) \right]
    \right\}' = 0 \,,
\ee
and for maximal symmetry in the $x^\mu$ directions and a symmetry
under shifts in $\theta$ this simplifies to
\be
    \Bigl[ \phi' \Bigr] + \frac{\kappa^2}{\sqrt{-\hg}} \,
    \frac{\delta S_b}{\delta \phi} = \Bigl[ \phi' \Bigr]
    - \frac{\kappa^2}{\sqrt{-\hg}}
    \left\{ \sqrt{-\hg} \left[
    \coneTensn  + \frac12 \,  Y_1 \cR \right]  \right\}' = 0 \,,
\ee

\section{Explicit Solution to the jump conditions when $R=0$}
\label{Appendix: solvingjumps}

We next explicitly solve the junction conditions in $D$ spacetime
dimensions for the special case of the flat, $R=0$, solutions
given in the main text, using only lowest-derivative terms in the
brane action. We therefore take the interior solution to be the
trivial one: constants $W_i = 0$ and $\phi_i = \phi_b$, and
$e^{B_i} = \rho$. The exterior solution by contrast is given by
\be
 e^{\phi_e} = e^{\phi_b} \left( \frac{\rho + \ell}{\rho_b + \ell}
 \right)^\gamma \,, \qquad
 e^{W_e} = \left( \frac{\rho + \ell}{\rho_b + \ell}
 \right)^\omega \qquad \hbox{and} \qquad
 e^{B_e} = \rho_b \left( \frac{\rho + \ell}{\rho_b + \ell}
 \right)^\beta \,,
\ee
where $\rho_b > -\ell$ and continuity of $\phi$, $W$ and $B$ from
the cap geometry to the exterior bulk have been used. The bulk
field equations imply the powers $\omega$, $\beta$ and $\gamma$
satisfy eq.~\pref{kasnerconditions}: $(D-2) \omega + \beta = (D-2)
\omega^2 + \beta^2 + \gamma^2 = 1$.

The derivative discontinuities at the brane are
\be
    \Bigl[ \partial_\rho \phi \Bigr]_b =
    \frac{\gamma}{\rho_b + \ell} \,, \quad
    \Bigl[ \partial_\rho W \Bigr]_b = \frac{\omega}{\rho_b + \ell}
    \quad\hbox{and}\quad
    \Bigl[ \partial_\rho B \Bigr]_b
    = \frac{\beta}{\rho_b + \ell} - \frac{1}{\rho_b}
    = \frac{(\beta-1)\rho_b - \ell}{\rho_b(\rho_b+\ell)}\,,
\ee
so the jump conditions become
\ba
    \Bigl[ \phi' \Bigr]_b &=& \frac{\gamma}{\rho_b + \ell}
    = \frac{\kappa^2}{\rho_b} \left\{ \rho_b \coneTensn (\phi_b)
    + \frac{n^2}{2\rho_b} \, \coneKin (\phi_b) \right\}' \nn\\
    \Bigl[ W' - B' \Bigr]_b &=&  \frac{\omega-\beta}{\rho_b + \ell}
    + \frac{1}{\rho_b} = \frac{(\omega - \beta + 1)\rho_b +
    \ell}{\rho_b (\rho_b + \ell)} = \frac{n^2 \kappa^2
    \coneKin  }{\rho_b^2 } \nn\\
    \Bigl[ W' \Bigr]_b &=&  \frac{\omega}{\rho_b + \ell}
    = - \frac{\kappa^2}{D-2} \, \left\{ \coneTensn  - \frac{
    n^2}{2 \rho_b^2 } \, \coneKin  \right\} \,.
\ea

We first solve for $\omega$, $\beta$ and $\gamma$, using the $W'$
jump condition together with eqs.~\pref{kasnerconditions}.
Defining
\be
    \X = (\rho_b + \ell) \kappa^2 \left\{ \coneTensn  -
    \frac{n^2}{2\rho_b^2}  \, \coneKin  \right\}\,,
\ee
we find
\be \label{app:powersolnv1a}
    \omega = - \frac{\X}{D-2} \,, \quad
    \beta = 1 + \X \quad \hbox{and}\quad
    \gamma^2 = -\X \left( 2 + \frac{D-1}{D-2} \, \X \right) \,.
\ee
As before, the condition $\gamma^2 \ge 0$ implies
\be
    - \frac{D-2}{D-1} \le \frac{\X}{2} \le 0 \,,
\ee
and so the condition $\rho_b > -\ell$ only allows solutions in the
right range to exist if $\coneTensn  < n^2 \coneKin /(2\rho_b^2)$.
This range for $\X$ also implies
\be \label{app:inequalities1}
    0 \le \omega \le \frac{2}{D-1} \,,
    \quad -\frac{D-3}{D-1} \le \beta \le 1
    \quad \hbox{and} \quad
    0 \le \gamma^2 \le \frac{D-2}{D-1} \,.
\ee

To eliminate $\ell$ use the $\phi'$ junction condition,
$\gamma/(\rho_b + \ell) = (\kappa^2/\rho_b) \Bigl\{\rho_b
\coneTensn  + n^2 \coneKin '/(2\rho_b) \Bigr\}'$, to write
\be \label{app:newXeqn}
    \X = (\rho_b + \ell) \kappa^2 \left\{ \coneTensn  -
    \frac{n^2}{2\rho_b^2} \, \coneKin  \right\}
    = \frac{\gamma [\rho_b \coneTensn  - n^2 \coneKin /(2\rho_b)
    ]}{[\rho_b \coneTensn  + n^2 \coneKin
    /(2\rho_b)]'}  = \frac{\gamma(\redoneTensn  - \redoneKin )}{\redoneTensn ' + \redoneKin '} \,,
\ee
where the dimensionless quantities $\redoneTensn  = \rho_b
\kappa^2 \coneTensn $ and $\redoneKin  = {n^2 \kappa^2 \coneKin
}/({2\rho_b})$ are as defined in the main text. Using
eq.~\pref{app:newXeqn} in the solution,
eq.~\pref{app:powersolnv1a}, allows $\gamma$ to be solved
completely in terms of $\rho_b$, $\coneTensn $ and its
derivatives. This leads to the expressions
\be \label{app:powersvschi1}
    \omega = \frac{2\,\chi^2}{(D-2) + (D-1) \chi^2} \,, \quad
    \beta = \frac{(D-2) - (D-3) \chi^2}{(D-2) + (D-1) \chi^2}
    \quad\hbox{and}\quad
    \gamma = - \frac{2(D-2)\chi}{(D-2) + (D-1)\chi^2} \,,
\ee
where now
\be \label{app:chi-inv-vs-UV}
    \frac{1}{\chi} = \frac{[\rho_b \coneTensn  + n^2 \coneKin
    /(2\rho_b)]'}{\rho_b \coneTensn  - n^2 \coneKin /(2\rho_b)}
    = \frac{\redoneTensn ' + \redoneKin '}{\redoneTensn  - \redoneKin }  \,.
\ee
Notice that these satisfy the inequalities,
eqs.~\pref{app:inequalities1}, for all $\chi$, with the additional
information that the signs of $\gamma$ and $\chi$ are opposite.

We solve for the ratio $\ell/\rho_b$ using the $[W'-B']$ junction
condition, in the form
\be \label{app:ratiovschi1}
    \omega -\beta + 1 + \frac{\ell}{\rho_b} =
    + \frac{(\rho_b+\ell) n^2 \kappa^2 \coneKin }{\rho_b^2}
    = \left( 1 + \frac{\ell}{\rho_b} \right) 2\, \redoneKin  \,,
\ee
and find
\be \label{app:ratiovschi1a}
    \frac{\ell}{\rho_b} = \frac{2\,\redoneKin  - (D-1)\omega}{1 - 2\,\redoneKin }
    = \frac{1}{1 - 2\,\redoneKin } \left\{ 2\,\redoneKin  -
    \frac{2(D-1)\,\chi^2}{(D-2) + (D-1) \chi^2}   \right\}  \,.
\ee
Notice that $\rho_b > -\ell$ implies $\ell/\rho_b > -1$.
Alternatively, we may solve for $\rho_b$ by instead using the
expression for $\omega$ as a function of $\X$, to get
\be
    (D-1)\omega
    = - \frac{D-1}{D-2}\, \X
    = -\frac{D-1}{D-2} \left(\frac{\ell}{\rho_b} +1\right) \kappa^2
    \left\{ \rho_b \coneTensn  - \frac{n^2 \coneKin }{2\rho_b} \right\}
    = \frac{D-1}{D-2} \left(\frac{\ell}{\rho_b} +1\right)
    \Bigl(\redoneKin  - \redoneTensn  \Bigr) \,,
\ee
so
\be \label{app:ratiovsY1}
    \frac{\ell}{\rho_b} +1 = \frac{(D-2) \omega}{\redoneKin  - \redoneTensn } \,.
\ee
Eliminating $\ell$, by combining eqs.~\pref{app:ratiovschi1a} and
\pref{app:ratiovsY1}, gives an expression involving only $\omega$,
$\redoneTensn $ and $\redoneKin $ (or, equivalently, only $\chi$,
$\redoneTensn $ and $\redoneKin $):
\be \label{app:rhomatching1}
    \frac{(D-1) \,\omega -1}{1 - 2\,\redoneKin } = \frac{(D-2)\omega}{\redoneTensn -\redoneKin }
    \,.
\ee

To get the final relation relating $\rho_b$ to $\phi_b$, eliminate
$\omega$ in terms of $\chi$ and use eq.~\pref{app:chi-inv-vs-UV}
to remove $\chi$, as in
\ba
    \frac{1}{\omega} &=& \frac{(D-1) \, \redoneTensn  + (D-3) \redoneKin  - (D-2)}{\redoneTensn  - \redoneKin }
    \nn\\
    &=&  \frac{D-1}{2} + \left(
    \frac{D-2}{2} \right) \frac{1}{\chi^2}
    = \frac{D-2}{2} + \frac{D-2}{2} \left[ \frac{(\redoneTensn ' + \redoneKin ')}{\redoneTensn -\redoneKin }
    \right]^2 \,,
\ea
where the first line follows from eq.~\pref{app:rhomatching1} and
the second line uses eqs.~\pref{app:powersvschi1} and
\pref{app:chi-inv-vs-UV}. This can be rewritten somewhat to give
the constraint
\be \label{app:newDEUV}
    (D-2) (\redoneTensn ' + \redoneKin ')^2 + (D-1) (\redoneTensn -\redoneKin )^2 + 2(\redoneTensn -\redoneKin )
    \Bigl[ (D-2) - (D-1)\,\redoneTensn  - (D-3)\redoneKin  \Bigr] = 0 \,.
\ee
Notice that this agrees with the appropriate specialization of
eq.~\pref{braneconstraint}: to $R=W_b=V=0$.

\medskip\noindent{\bf Conical bulk solution}

\medskip\noindent The special case where the external geometry is
a cone corresponds to the choices $\omega = \gamma = 0$ and $\beta
= 1$. As the $W'$ junction condition shows, this is only possible
(given a flat cap geometry) if $t_{\theta\theta} = 0$, and so
\be \label{app:conecondition}
    \rho_b \coneTensn  = \frac{n^2 \coneKin }{2\rho_b} \quad\hbox{or}
    \quad \redoneTensn  = \redoneKin  \,.
\ee
Used in the above formulae this implies $\chi =0$, which ensures
the vanishing of both $\omega$ and $\gamma$, as claimed.
Eq.~\pref{app:conecondition}, implies $\rho_b$ is given by
\be \label{app:naiverhocone}
    \rho_b^2 = \frac{n^2 \coneKin }{2 \coneTensn } \,,
\ee
which when used in the definition of $\redoneTensn $ implies
\be \label{app:naiverhocone1}
    \redoneTensn  = \kappa^2 \rho_b \coneTensn  = \kappa^2 \sqrt{ \frac{n^2 \coneTensn
    \coneKin }{2}}\,.
\ee
Consequently,
\be
    \frac{\ell}{\rho_b} = \frac{2\,\redoneKin }{1- 2\,\redoneKin }
    \,.
\ee

Finally, the condition that $\phi'$ must vanish at the brane
(recall $\gamma = 0$) requires $\redoneTensn +\redoneKin  =
2\redoneTensn $ to be $\phi$-independent, but this is only
consistent with eq.~\pref{app:naiverhocone1} if the product
$\coneTensn  \coneKin $ is independent of $\phi$.

If, for example, we take $\coneTensn (\phi) = U_0 \, e^{u \phi}$
and $\coneKin (\phi) = V_0 \, e^{v \phi}$, then we may specialize
the above conical-limit equations to
\be
    \rho_b = \sqrt{\frac{n^2 V_0}{2U_0}} \; e^{(v-u)\phi_b/2}
    \quad\hbox{and} \quad
    \redoneTensn  = \redoneKin  = \kappa^2 \rho_b \coneTensn
    = \kappa^2 \sqrt{\frac{n^2 U_0 V_0}{2}} \; e^{(v+u)\phi_b/2}
    \,,
\ee
which show that $\redoneTensn $ is only $\phi$-independent if $u =
-v$. Notice that this includes in particular the case $u = -v =
2/(D-2)$ which encodes scale invariance in the bulk field
equations. Notice also that the choice $u = -v$ also implies
$\rho_b \propto e^{v\phi_b}$, and so is $\phi_b$-dependent unless
$u=v=0$.

\end{document}